\documentclass[apj,numberedappendix]{emulateapj_051214}
\journalinfo{\rightline {To appear in The Astrophysical Journal}}
\submitted{Received 2005 August 8; accepted 2005 December 12}

\newcommand{\etal}{et al.}

\shorttitle{Nonlinear Decline-Rate Dependence of SNe Ia}
\shortauthors{Wang, Strovink {et al.}}

\begin{document}

\title{Nonlinear Decline-Rate Dependence and Intrinsic Variation \\
       of Type I{\lowercase{a}}\ Supernova Luminosities}

\author{Lifan Wang,\altaffilmark{1,3} Mark Strovink,\altaffilmark{1,2} Alexander Conley,\altaffilmark{1,2,4} Gerson Goldhaber,\altaffilmark{1,2}\\
Marek Kowalski,\altaffilmark{1} Saul Perlmutter,\altaffilmark{1,2} and James Siegrist\altaffilmark{1,2}} 

\affil{$^1$E.~O.~Lawrence Berkeley National Laboratory, Berkeley, CA 94720}
\affil{$^2$Physics Department, University of California, Berkeley, CA 94720}
\affil{$^3$Purple Mountain Observatory, Nanjing 210008, China}
\email{strovink@lbl.gov}

\altaffiltext{4}{Now at Department of Astronomy and Astrophysics, University of Toronto, Ontario, Canada M5S 3H8}

\begin{abstract}
Published $B$ and $V$ fluxes from nearby Type Ia supernov\ae\ are fitted to light-curve templates with 4-6 adjustable parameters.  Separately, $B$ magnitudes from the same sample are fitted to a linear dependence on $\bv$ color within a post-maximum time window prescribed by the {\sc cmagic} method.  These fits yield two independent SN magnitude estimates $B_{\rm max}$ and $B_{BV}$.  Their difference varies systematically with decline rate $\Delta m_{15}$ in a form that is compatible with a bilinear but not a linear dependence; a nonlinear form likely describes the decline-rate dependence of $B_{\rm max}$ itself.  A Hubble fit to the average of $B_{\rm max}$ and $B_{BV}$ requires a systematic correction for observed $\bv$ color that can be described by a linear coefficient ${\cal R} = 2.59 \pm 0.24$, well below the coefficient $R_B \approx 4.1$ commonly used to characterize the effects of Milky Way dust.  At 99.9\% confidence the data reject a simple model in which no color correction is required for SNe that are clustered at the blue end of their observed color distribution.  After systematic corrections are performed, $B_{\rm max}$ and $B_{BV}$ exhibit mutual rms intrinsic variation equal to $0.074 \pm 0.019$ {mag}, of which at least an equal share likely belongs to $B_{BV}$.  SN magnitudes measured using maximum-luminosity or {\sc cmagic} methods show comparable rms deviations of order $\approx$0.14 {mag} from the Hubble line.  The same fit also establishes a 95\% confidence upper limit of 486 km s$^{-1}$ on the rms peculiar velocity of nearby SNe relative to the Hubble flow.
\end{abstract}

\keywords{supernov\ae: general --- cosmology: observations --- distance scale}

\section{INTRODUCTION} \label{intro}

Well before the Universe was found to be accelerating \citep{Perlmutter:1998, Garnavich:1998, Schmidt:1998, Riess:1998, Perlmutter:1999}, Type Ia supernov\ae\ (SNe), the study of which underpinned that discovery, were thought to be standard candles.  Even before the Type Ia subclass was identified, \citet{Kowal:1968} studied the distribution of Type I SN luminosities, and \citet{Pskovskii:1977} observed that brighter SNe exhibit slower decline rates.  The standardization of Type Ia SNe sharpened when \citet{Phillips:1993} introduced a brighter-slower correction that was linear in the decline-rate parameter $\Delta m_{15}$ (the change in magnitude from maximum light to 15 rest-frame days thereafter).  That correction was refined by \citet{Hamuy:1996a} and by \citet{Phillips:1999} (henceforth Phil99), who added a quadratic term.  \citet{Tripp:1998}, \citet{Tripp:1999}, and \citet{Parodi:2000} fit a correction that was linear both in $\Delta m_{15}$ and in a second parameter, $\bv$ color.  In these two-parameter studies the observed brighter-bluer correlation was compensated by a fit color coefficient ${\cal R} \approx 2.5\,$.  \citet{Perlmutter:1997} and \citet{Goldhaber:2001} (henceforth Perl97 and Gold01) stretched the time axis of a fixed template to approximate different SN light curves; that linear stretch factor substituted for $\Delta m_{15}$ in their alternative brighter-slower correction.

Given an adequate training set -- here an ensemble of SNe in the nearby Hubble flow whose deviations $\Delta_i$ from a fiducial absolute SN magnitude are deduced from their redshifts -- the problem of estimating the deviation $\Delta$ of a single SN, using a set $\{x_j\}$ of its measured parameters, is amenable to general solution.  In principle the $x_j$ can include any parameter that may add to knowledge of $\Delta$: fluxes in various bands at various phases, spectral line amplitudes, {\it etc}.  One example of such a solution is given in Appendix \ref{appS}.  Present progress along these more general lines is exemplified by the {\sc mlcs}2k2 package and its predecessors described by \citet{Riess:1995,Riess:1996}, \citet{Riess:1998}, and \citet{Jha:2002}.  As those authors found, multivariate approaches to SN standardization are severely limited by low training statistics.  For example, the {\sc mlcs}2k2 training set did not adequately constrain the color-correction coefficient $R_V$, for which an external value was imposed.

This paper aims to extend understanding of the systematic corrections that help standardize Type Ia supernova luminosities, and of the intrinsic variations which blur that standard.  Large resources now are devoted to projects that use standardized SNe to measure with exquisite precision the parameters determining the Universe's recent acceleration, and even larger efforts will be needed to track the evolution of its equation of state, revealing the sources of that acceleration.  Even small advances in such understanding can have big implications for the realization of these projects.  

In this paper, we make use of only a few properties of nearby SNe -- luminosity, color and decline rate -- that are simple to characterize.  In that sense our approach to SN standardization is traditional, in the vein of work cited in the first paragraph.  Nevertheless, we do bring fresh tools to this study.  By taking the difference between the two simplest measures of SN magnitude -- luminosity at peak, and luminosity at fixed color using the {\sc cmagic} method \citep[][henceforth Wang03]{Wang:2003} -- we impose a new, more precise constraint on the (nonlinear) form of their decline-rate dependence, and we set a robust lower bound on the intrinsic variation  suffered by at least one of these measures.  By applying a newly extended light-curve fitting method and standard error propagation consistently to each SN in the sample, we obtain a Hubble fit that sets a significant upper bound on the rms peculiar velocity of SNe relative to the Hubble flow.

In Section \ref{fits} our light-curve ({\sc lc}) and {\sc cmagic} fitting methods are introduced.  Their outputs are shown to be independent, and the corrections made to them for extinction and time dilation are described.  Section \ref{fitPhoto} identifies the photometric data sample that is processed by these methods, and the subsample of SNe to which Hubble fits are applied.  Section \ref{DeltaE} reveals the new information gained by comparing the {\sc lc} and {\sc cmagic} fit magnitudes.  In Section \ref{Bavg} the result of a Hubble fit to the average of these outputs is presented; this is the final step in a global fit to the sample.  Section \ref{Bmax} uses the global fit parameters to quantify the deviations of the {\sc lc} and {\sc cmagic} outputs from the Hubble line.  Section \ref{blue} explores properties of the subset of SNe assumed to be largely free of extinction by host-galactic dust.  The paper concludes with a summary and discussion. 
 
\section{METHODS FOR FITS TO PHOTOMETRIC DATA} \label{fits}

We used two complementary methods to estimate SN magnitudes.  The first, a light-curve method described in \S \ref{LCfits}, is based on the fit peak magnitudes in the $B$ and $V$ bands, and on the fit difference $\Delta m_{15}$ between the peak $B$ magnitude and its value 15 rest-frame days later.  These three key numbers and their uncertainties are supplied by our own {\sc lc} fits to published measured fluxes in these bands.  The fits were performed by a newly and substantially extended version of the algorithm of Perl97 and Gold01, in which the time axes of fixed templates were linearly stretched to best represent the flux data.  After color and decline-rate corrections, the SN magnitude yielded by our {\sc lc} method is called $B_{\rm max}$.

Our second method for estimating SN magnitudes, summarized in \S \ref{CMfits}, is based on the output of the {\sc cmagic} algorithm introduced by Wang03.  This algorithm exploits the nearly linear dependence of $B$ magnitude upon $\bv$ color over a period extending, for typical SNe, from $\approx$6 to $\approx$27 rest-frame days after maximum light.  A least-squares linear fit to published magnitudes and colors measured within that range yields an intercept $B_{BV}$ at fixed color $\bv = 0.6$.  The above mentioned {\sc lc} fit in the $B$ band is also used to supply $\Delta m_{15}$ and the phase at maximum.  After color and decline-rate corrections, the SN magnitude yielded by our {\sc cmagic} method is called $B_{BV}$.

By no means are either of these methods restricted to the $B$ and $V$ bands.  Much study has been devoted to their extension to the $R$, $I$, and $U$ bands (Wang03; Wang, L.\ \etal\ 2005, in preparation, henceforth Wang05a; Goldhaber, G.\ \etal\ 2005, in preparation).  However, since this paper relies on comparing the results of these two methods, sufficient data must exist to complete both types of fit for each SN.  Were we additionally to require successful {\sc lc} and {\sc cmagic} fits to the same SNe in the $R$, $I$, or $U$ band, our sample size at best would halve.  Data in these additional bands therefore remain outside this paper's scope.
 
\subsection{Light Curve Fits} \label{LCfits}

This study requires use of a light-curve fit algorithm only for interpolating or extrapolating measured fluxes to compute $B_{\rm max}$, $V_{\rm max}$, and $\Delta m_{15}$.  Here no other use is made of the {\sc lc} fit parameters, and no physical significance need be ascribed to them.  For completeness, neverthess, we describe briefly the parametrization that we employed.  More information on this algorithm is supplied in Wang05a.

The set of published flux measurements (for example in the $B$ band) for a single SN was fit to the function
\begin{eqnarray}
f(t;\Gamma_0,t_0,s_0,\tau_0,&\epsilon_0&\!,t_d) = \Gamma_0 [{\cal T}_B(t^*(t-t_0;s_0,\tau_0)) + \nonumber \\
+ \, &\epsilon_0& {\cal T}_B(t^*(t-t_0-t_d;s_0,\tau_0))] \label{eqc}
\end{eqnarray}
where $t$ is the rest-frame phase of the SN flux measurement; $\Gamma_0$ is an overall normalization; ${\cal T}_B(t_r)$ is a fiducial template for $B$ band flux {\it vs.}~rest-frame phase $t_r$ that peaks at $t_r = 0$; and $t_0$ is the phase at which the first (main) term reaches its peak.  In the first term, the ``stretched'' phase $t^*$, discussed further in Appendix \ref{appLC}, is an odd function of $t-t_0$ with parameters $s_0$ and $\tau_0$.  The ``stretch'' parameter $s_0$ is equal to ${\rm d}t/{\rm d}t^*$ at $t=t_0$; the ``duration'' parameter $\tau_0$ is such that ${\rm d}t/{\rm d}t^*$ is damped to unity when $|t-t_0| \gg \tau_0$.  Relative to the main term, the second (delayed) term is different only in that its phase at peak amplitude is delayed by $t_d$ and its amplitude is reduced by the factor $\epsilon_0$.  The special case \{$\tau_0 = \infty$ (no damping); $\epsilon_0 = 0$ (no delayed term)\} was studied by Perl97 and Gold01; under those conditions their parameter $s$ is equivalent to $s_0$.  In all fits here, the parameter $\tau_0$ is allowed to vary along with $\Gamma_0$, $t_0$, and $s_0$; when $\epsilon_0$ is nonzero, as is permitted (Appendix \ref{appLC}) for a minority of SNe, $\epsilon_0$ and $t_d$ vary as well.  Within a band, usually all available data were fitted, including those in the ``tail'' region $t^* \ga 30$ dy.  Here we are not concerned with results that are specific to the tail region, or the transition thereto, because we emphasize methods that may profitably be applied to faint SNe that (unlike the nearby sample) are sensitive to cosmological acceleration.

For neighboring bands the {\sc lc} fit procedure is the same (except for choice of template).  Smoothed versions of the $B$ template described by Gold01 and of the $V$ template described by \citet{Knop:2003} are used.  Parameters that are fit to data in different bands are nearly independent, with only two exceptions:  for all bands in which a given SN is fitted, $\epsilon_0$ either is left free or is set to zero; and the epochs of peak flux are required to align within certain loose windows.  To satisfy the latter conditions, all bands were fit at once.  For decline-rate corrections we did not use $\Delta m^V_{15}$ in the $V$ band as a substitute for, or in combination with, the $B$ band value: though a satisfactory correlation of $\Delta m^V_{15}$ with $\Delta m_{15}$ was seen, and reasonable Hubble fits using $\Delta m^V_{15}$ alone or in combination with $\Delta m_{15}$ were produced, tighter Hubble residuals were obtained by using $\Delta m_{15}$ alone.  Neither did we substitute the fit parameter $1/s_0$ for $\Delta m_{15}$, because $\Delta m_{15}$ is correlated also with the companion parameter $\tau_0$.

For various choices of the stretch function $t^*(t)$ in equation (\ref{eqc}), our {\sc lc} fit results for $B_{\rm max}$ and $V_{\rm max}$ are stable.  Measuring $\Delta m_{15}$ is more delicate: because the time derivative of the flux at day 15 can be large, a small error in the definition of the date of maximum flux can translate into a large error in $\Delta m_{15}$.  Especially near those times, both $t^*(t)$ and the template parametrization must be smoothly differentiable.  Accordingly we employed a particular choice of $t^*(t)$ for which details are supplied in Appendix \ref{appLC}.  

In summary, we have applied to the full SN sample a uniform mathematical approach for determining their peak magnitudes and decline rates.  The approach is self-consistent, with covariance matrices used to propagate the raw flux errors through to the final uncertainties.  As a check of these methods, we subjected accurate graphs of each of our {\sc lc} fits to detailed inspection which confirmed that, within quoted uncertainties, the measured fluxes are represented satisfactorily by these fits.  

As an alternative approach, we could have attempted to combine and use the published results of different light-curve fits, employing various methodologies, that had been performed on these SNe.  The dangers in such an attempt become evident when one compares our values of parameters such as $\Delta m_{15}$ and peak magnitude with those found {\it e.g.}\ in the useful compilation of \citet{Reindl:2005}.  Significant differences are found, particularly in values of $\Delta m_{15}$.  For example, the compiled values of $\Delta m_{15}$ are largely confined above a lower limit of $\approx 0.84$-$0.87$ mag, with several exactly equal to 0.87.  Other SNe are clustered at $\Delta m_{15} = 1.13$ and 1.69.  Such artifacts, which are not found in our own $\Delta m_{15}$ distributions, appear to have arisen from limitations in the number of well-observed SN light curves that were available for use as discrete templates by the published fits.    

\subsection{Fits to $B$ Magnitude {\it vs.}~$\bv$ Color} \label{CMfits}

The {\sc cmagic} method described by Wang03 begins with pairs of fluxes measured in the $B$ and $V$ bands at the same epoch $t_i$.  These flux pairs are converted to raw magnitude pairs $\{B_i,V_i\}$ and entered on a plot of $B_i$ {\it vs.}~$(B_i-V_i)$.  In parallel, the above-described {\sc lc} fit is used to determine the decline-rate parameter $\Delta m_{15}$ and the rest-frame phase $t_{\rm peak}$ of peak $B$ flux.  To those pairs $\{B_i,V_i\}$ which lie within a fiducial region $t_b - t_{\rm peak} < t_i < t_e - t_{\rm peak}$, a weighted least-squares linear fit is performed in which the errors on $B_i$ and $V_i$ are taken to be independent.  At least three pairs must participate in the fit, providing a minimum of one constraint.  The fit outputs are $B^{\rm raw}_{BV}$, the intercept of the best-fit line with the fixed color $B_i-V_i = 0.6$ {mag}; the slope $\beta_{BV}$ of that line; and their errors.   (Our $B^{\rm raw}_{BV}$ is equivalent to $B_{BV0.6}$ of Wang03 and \citet{Conley:2005} (henceforth Conl05).)  The boundaries $\{t_b-t_{\rm peak},t_e-t_{\rm peak}\}$ of the {\sc cmagic} linear region are $\{ (7 \; {\rm mag})/\Delta m_{15}, (30 \; {\rm mag})/\Delta m_{15} \}$ dy for normal SNe, and $\{ (10 \; {\rm mag})/\Delta m_{15}, (29 \; {\rm mag})/\Delta m_{15} \}$ dy for spectroscopically SN1991T-like SNe.

As noted by Wang03, the average slope $\beta_{BV}$ is very nearly equal to 2, with an rms intrinsic scatter (apart from photometric error) that is less than 0.1.  If $\beta_{BV}$ either were assumed or were measured always to have the same value, as is nearly the case, our choice of intercept color $B_i-V_i = 0.6$ would be immaterial.  Here this particular value is chosen for compatibility with earlier work (Wang03; Conl05), and for its consistency with the intercept color that minimizes both the measured rms Hubble residual of $B_{BV}$ and the $|$covariance$|$ between the intercept and $\beta_{BV}$.
\subsection{Photometric Errors} \label{scaling}

Sources of photometric SN data are diverse; reported errors on data points are not always reliable, and their covariances normally are unavailable.  Additional systematic errors arise from observations on different telescopes and from uncertainties in the adopted light curve models.  To gain confidence in the errors on {\sc lc} and {\sc cmagic} fit outputs, we examined the distributions of confidence levels ({\sc cl}s) for each type of fit.  Ideally both distributions should be uniform between 0 and 1.  However, typically the {\sc lc} fits had smaller {\sc cl}s than expected, likely because, even with 4-6 adjustable parameters, the templates were not flexible enough to fully track the detailed evolution of well-measured flux data.  Conversely, the {\sc cl}s for {\sc cmagic} fits typically were larger than expected, perhaps because published photometric errors in the {\sc cmagic} linear region are slightly conservative.  To address this issue for each type of fit, we mapped the observed {\sc cl} distribution onto a uniform one while maintaining each SN's rank in the distribution.  This was done by scaling all fit errors that were output for a particular SN by $(\chi^2_{\rm orig}/\chi^2_{\rm unif})^{1/2}$, where $\chi^2_{\rm orig}$ is the original $\chi^2$ statistic and $\chi^2_{\rm unif}$ is the value which allows that SN to maintain its rank in the uniform {\sc cl} distribution.  For the {\sc lc} and {\sc cmagic} fits, the median error-scaling factors were 1.57 and 0.78, respectively.
\subsection{Extinction of $B_{\rm max}$ and $B_{BV}$} \label{extinct}

As a first step, standard corrections for the effects of Milky Way ({\sc mw}) dust were applied to all of the {\sc lc} and {\sc cmagic} fit outputs that require it.  This step will be discussed in \S \ref{fitPhoto}.  In this section, on the other hand, we are concerned with extinction by intervening dust outside the {\sc mw}.  An experimentally related issue is the effect upon SN luminosity of variation in intrinsic SN color.

If intervening dust outside the {\sc mw} causes true $\bv$ to differ from raw ({\it i.e.}~observed) $(\bv)^{\rm raw}$ by $\delta(\bv) \equiv (\bv) \!-\! (\bv)^{\rm raw}$, then, according to the law of \citet{Cardelli:1989} (henceforth Card89), all true $B$ magnitudes differ from raw $B$ magnitudes by $\delta B \equiv B \!-\! B^{\rm raw} = R_B \, \delta(\bv)$, where $R_B$ is approximately constant and of size $\approx 4.1$ if the dust is parametrized as is common for {\sc mw} dust.  For maximum-luminosity analysis, the SN color is measured here by the difference $E \equiv B^{\rm raw}_{\rm max} - V^{\rm raw}_{\rm max}$ between the $B$ magnitude at $B_{\rm max}$ and the $V$ magnitude at $V_{\rm max}$, as determined, respectively, by the nearly independent {\sc lc} fits in the $B$ and $V$ bands.  More generally, using the Card89 law to describe the effects on raw SN luminosity both of intrinsic SN color variation and of intervening dust, and, following \citet{van den Bergh:1995}, denoting by ${\cal R}$ the color-correction coefficient parametrizing those combined effects, color-corrected $B_{\rm max}$ is related to color-uncorrected $B^{\rm raw}_{\rm max}$ by
\begin{equation}
B_{\rm max} = B^{\rm raw}_{\rm max} - {\cal R}E \; . \label{eqC}
\end{equation}
(Beginning in \S \ref{DeltaE}, we shall use the symbol $B_{\rm max}$ to denote the maximum $B$ magnitude after full corrections, which depend both on observed color and on decline rate, are applied.  In the present discussion, nevertheless, for simplicity we do not include a decline-rate correction to $B^{\rm raw}_{\rm max}$.) 

For {\sc cmagic} analysis, however, equation (\ref{eqC}) does not hold: the change due to intervening dust in the {\sc cmagic} output $B_{BV}$, defined as the $B$ magnitude at fixed color $\bv$, will be smaller.  Because it must continue to be measured at the same fixed color, the true value of $B_{BV}$ should differ from the raw value $B^{\rm raw}_{BV}$ by only that portion of $\delta B$ which is orthogonal to $\delta(\bv)$:
\begin{eqnarray}
\delta B_{BV} &=& \delta B - {{{\rm d}B} \over {{\rm d}(\bv)}} \; \delta(\bv)
  \nonumber \\
&=& (R_B - \beta_{BV})\, \delta (\bv) \; . \label{eqA}
\end{eqnarray}
Again taking into account the effects both of intrinsic SN color variation and of intervening dust, the corresponding effective color-correction coefficient for $B_{BV}$ is ${\cal R} - \beta_{BV}$.

In raw form, before correction for color or decline rate, $B^{\rm raw}_{BV}$ uses only the $B$ and $V$ fluxes that are measured in the {\sc cmagic} linear region, beginning at least several days after maximum light.  Nevertheless, to define that region and to measure the decline rate, in parallel one performs a light-curve fit that does also use the fluxes measured near maximum, normally in the $B$ band.  To measure and correct for the SN color, in principle a second {\sc lc} fit in the $V$ band would allow the maximum-luminosity color $E$ to be calculated.  However, to preserve greatest orthogonality to maximum-luminosity analysis, the {\sc cmagic} method substitutes a complementary measure of SN color that is unaffected by the details and uncertainties of the $V$ light curve near maximum.  This is done by exploiting the unique linear combination
\begin{equation}
{\cal E} \equiv {{B^{\rm raw}_{\rm max} - B^{\rm raw}_{BV}} \over {\beta_{BV}}} \label{eqB}
\end{equation} 
of $B^{\rm raw}_{\rm max}$ and $B^{\rm raw}_{BV}$ that permits $\delta {\cal E} = \delta(\bv)$ in the example to which equation (\ref{eqA}) pertains.  For {\sc cmagic} analysis, ${\cal E}$ is not the only possible color measure, but it is a choice that has become conventional and is adopted throughout this paper.

In analogy to equation (\ref{eqC}), the arguments following equation (\ref{eqA}) require 
\begin{equation}
B_{BV} = B^{\rm raw}_{BV} - ({\cal R} - \beta_{BV}){\cal E} \; . \label{eqD}
\end{equation} 
Using equation (\ref{eqB}) to evaluate ${\cal E}$, equation (\ref{eqD}) becomes
\begin{equation}
B_{BV} = B^{\rm raw}_{\rm max} - {\cal R}{\cal E} \; . \label{eqE}
\end{equation} 
Comparison of eqs.~({\ref{eqE}) and (\ref{eqC}) emphasizes that $B_{BV}$ and $B_{\rm max}$ are independent only in that they use different measures of SN color to correct $B^{\rm raw}_{\rm max}$.  When the {\sc cmagic} color ${\cal E}$ closely tracks the maximum-luminosity color $E$, as is usually the case, $B_{BV}$ and $B_{\rm max}$ maintain the same sensitivity to uncertainties in ${\cal R}$.    
\subsection{Covariance of $B_{\rm max}$ and $B_{BV}$} \label{nondeg}

Given the close relationship between $B_{\rm max}$ and $B_{BV}$ revealed by eqs.~({\ref{eqC}) and (\ref{eqE}), it is natural to examine their mutual correlation.  Using a Monte Carlo technique, Conl05 studied the covariance due to photometric error between $B^{\rm raw}_{BV}$ and $B^{\rm raw}_{\rm max}$ after correcting these quantities only for decline-rate dependence.  In that study, $B^{\rm raw}_{\rm max}$ was determined by an {\sc lc} fit of the type used by Perl97 and Gold01, as discussed in \S \ref{LCfits}.  A mean Pearson correlation coefficient of $\approx$0.15 was determined.  Since a covariance at that level would not have had a significant effect, correlations due to photometric error between entirely uncorrected $B^{\rm raw}_{BV}$ and $B^{\rm raw}_{\rm max}$ were not taken into account in the error propagation performed here.  On the other hand, correlations due to photometric error in {\it corrections} to $B_{BV}$ and $B_{\rm max}$ are more significant: for example, after extinction corrections, both quantities depend on $B^{\rm raw}_{\rm max}$.  Such correlations are included straightforwardly in our error propagation, as are correlations due to peculiar velocity and extinction by {\sc mw} dust.

On a related note, it is of possible academic interest to identify the conditions, if any, that would cause corrected $B_{BV}$ and $B_{\rm max}$ to become fully correlated.  We find that when $\beta_{BV}$ is fixed, and stretch and time-of-maximum are determined for both $B$ and $V$ bands by a $B$-template fit, the maximum-luminosity and {\sc cmagic} methods are fully correlated if ``magic templates'' are used and if $V$ observations outside the {\sc cmagic} linear region are discarded; and if the number of $B$ observations is minimal, or if the $B$ points used by {\sc cmagic} are taken from the $B$ template fit.  (``Magic templates'' $m_B$ and $m_V$ satisfy ${\rm d}m_B/{\rm d}(m_B \!-\! m_V) = \beta_{BV}$ in the {\sc cmagic} linear region.)  None of these conditions are met by the analysis reported here.  Appendix \ref{appD} demonstrates that $B_{BV}$ and $B_{\rm max}$ indeed are fully correlated for a hypothetical case that does satisfy these requirements.

\subsection{$K$ Corrections} \label{Kcorrec}

The Hubble expansion not only dilates the apparent SN evolution, but also causes its light to be bluer at rest-frame emission than at detection.  To translate published photometric data from observed to standard rest-frame passbands, we employed a particular $K$-correction method, discussed more fully in Wang05a, that is described only briefly below.

For the actual range of redshifts $0.003 < z < 0.079$ of the SNe in our fitted sample, light detected at the peak of the $V$ ($B$) passband was emitted at a rest-frame wavelength that is blueshifted by up to 35\% (60\%) of the wavelength interval by which the peak of the adjacent $B$ ($U$) passband is separated.  Nevertheless, in part because the light curve is fit nearly independently in each passband, the functional forms used by both our {\sc lc} (\S \ref{LCfits}) and {\sc cmagic} (\S \ref{CMfits}) fits are flexible enough to permit both types of fit to be performed to the {\it observed} data {\it before} $K$ corrections for these wavelength shifts were applied; a posterior $K$ correction needed to be applied only to the five fit outputs $B^{\rm raw}_{\rm max}$, $V^{\rm raw}_{\rm max}$, $\Delta m_{15}$, $B^{\rm raw}_{BV}$, and $\beta_{BV}$.  We index these outputs by $k$ (1$\le$$k$$\le$5).

The framework for our $K$ corrections begins with that established by \citet{Hamuy:1993}, who directly applied the spectrophotometric data from three library SNe to the task of $K$-correcting a time series of observed $B$ and $V$ magnitudes belonging to other SNe that lacked spectral data.  At present we benefit from a library of $>$30 SNe for which spectroscopic as well as good photometric data are available.  For the $B$ and $V$ bands we constructed a set of SN light curve pairs, with each pair characterized by a different raw color $E_i \, $.  We then used the spectroscopic data to modify each pair so that it corresponded to each of a set of hypothetical redshifts $z_j$.  After applying standard {\sc lc} and {\sc cmagic} fit procedures to the entire collection of light curve pairs, we recorded the differences $\{K_k(E_i,z_j)\}$ between fit output $k$ at $z=z_j$ and the same output at $z=0$.  For a given $j$ and $k$, the corrections $\{K_k(E_i,z_j)\}$ were fit to a linear function of the $E_i \, $, with parameters $\mu^k_j$ and $\psi^k_j$.  In turn, the $\{\mu^k_{j}\}$ and $\{\psi^k_{j}\}$ each were fit to cubic polynomial functions of the $z_j$.  When applied to a real SN, the $K$ correction to the $k^{\rm th}$ fit output was calculated by evaluating the polynomial that was fit to the $\{K_k(E_i,z_j)\}$ at that SN's raw color $E$ and redshift $z$.

We found this scheme attractive because its implementation is straightforward and because the $K$ corrections are smooth functions of $z$ and, like $B_i$ itself in the {\sc cmagic} region, they are linear functions of color.  Examples of results from this scheme are provided in Appendix \ref{appK}.            

\section{FITS TO PHOTOMETRIC DATA} \label{fitPhoto}

For inclusion in this analysis, the available sample of nearby SNe was subjected to a set of standard cuts that are listed in Table \ref{cut}.  First, as noted in \S \ref{fits}, enough data were needed so that both a light-curve and {\sc cmagic} fit could be made to each SN.  Primarily this meant that at least three pairs $\{B_i,V_i\}$ of magnitudes were required to have been measured in the {\sc cmagic} linear region.  These requirements eliminated one quarter of the candidates.  Secondly, to avoid sensitivity to cosmological acceleration and to bound the range of redshift over which $K$ corrections are performed (\S \ref{Kcorrec}), we required $z < 0.1$.  Thirdly, we consider highly subluminous and fast-declining ``SN1991bg-like'' SNe to be a recognizably distinct subset.  If lumped together with the main sample, they would only muddy its characteristics.  As this $\approx$6\% subset is too small to analyze separately, we simply cut it out by requiring $\Delta m_{15} < 1.7$ {mag}.  Fourthly, a standard loose cut $E < 0.5$ {mag} on raw $\bv$ color eliminated only two severely reddened candidates.  As noted in \S \ref{CMfits}, the {\sc cmagic} fit slope $\beta_{BV}$ is clustered near 2 with an intrinsic scatter $<$0.1.  Our fifth cut, $1.5 < \beta_{BV} < 2.5$, eliminated four SNe whose {\sc cmagic} slopes are pathological and/or poorly measured.  Finally we discarded a single candidate, SN2000cx, whose properties are widely considered to be highly atypical of SNe Ia.  (Other candidates that also might be considered to be peculiar failed earlier cuts.)  These standard cuts left a ``base sample'' of 70 SNe.

\tabletypesize{\footnotesize}
\setlength{\tabcolsep}{2.1pt}
\begin{deluxetable}{lcc}
\tablewidth{0pt}
\tablecolumns{3}
\tableheadfrac{}
\tablecaption{Number of SNe passing cumulative cuts that yield the base, plotted, and fitted samples.\label{cut}}
\tabletypesize{}
\tablehead{\colhead{Condition satisfied (cumulative)}
&\colhead{\begin{tabular}{c} No.~of \\ SNe \\ retained \end{tabular}}
&\colhead{\begin{tabular}{c} Sample \\ name \end{tabular}}
}
\startdata
No.~of SNe initially considered&110&\nodata\\
Enough data to fit $B_{\rm max}$ and $B_{BV}$&\phn84&\nodata\\
$z < 0.1$&\phn82&\nodata\\
$\Delta m_{15} < 1.7$&\phn77&\nodata\\
$B_{\rm max} - V_{\rm max} < 0.5$&\phn75&\nodata\\
$1.5 < \beta_{BV} < 2.5$&\phn71&\nodata\\
Not peculiar (SN2002cx)&\phn70&base\\
$\sqrt{\sigma^2(B^{\rm raw}_{\rm max}) \!+\! 4\sigma^2(V^{\rm raw}_{\rm max}) \!+\! \sigma^2(B^{\rm raw}_{BV})} \!<\! 0.25$&\phn61&plotted\\
Each SN adds $\Delta \chi^2 < 8$ to global fit&\phn56&fitted\\
\enddata
\end{deluxetable}


To the base sample we applied a ``well-measured'' cut which requires explanation.  Normally, when several measurements of a physical quantity are combined and a proper error analysis is applied, nothing is lost by including measurements that are far less precise than those which dominate the result.  One of our main interests here, however, is the intrinsic variation of absolute SN magnitudes.  In the simplest picture, this is determined by subtracting the calculated variance due to photometric error from the observed total variance.  If unusually poorly measured SNe are included, both variances inflate, and the precision of their difference suffers.  As will be seen in \S \ref{declineDeltaE}, our main information on intrinsic SN variation comes from comparing the maximum-luminosity $\bv$ color $E$ to the {\sc cmagic} color ${\cal E}$.  As is discussed in Appendix \ref{appLC}, the photometric error on $\Delta E \equiv {\cal E} - E$ is tracked roughly by the simple quantity 
\begin{equation}
\Sigma \equiv \sqrt{\sigma^2(B^{\rm raw}_{\rm max}) + 4 \sigma^2(V^{\rm raw}_{\rm max}) + \sigma^2(B^{\rm raw}_{BV})} \; , \label{eqEF}
\end{equation}
where the $\sigma$'s are raw errors output by the {\sc lc} and {\sc cmagic} fit procedures described in \S \ref{LCfits} and \S \ref{CMfits}.  To identify a stable sample for further analysis, we chose to cut on $\Sigma$, as opposed to the precise error on $\Delta E$, because the latter depends on fit parameters that were expected to evolve as the analysis matured.   Figure \ref{sigdB} exhibits the distribution in $\Sigma$ of the 70 SNe in the base sample.  Remaining after the cut $\Sigma < 0.25$ are the 61 SNe which comprise the ``plotted sample''.


\begin{figure}
\epsscale{0.88}
\plotone{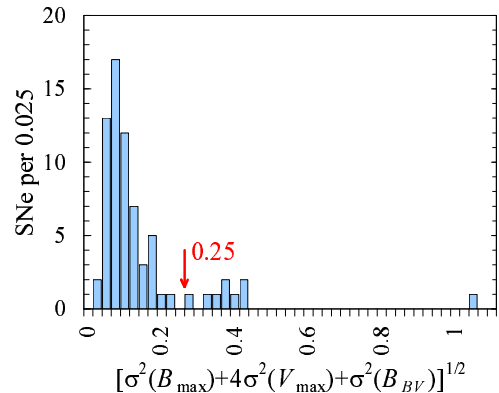}
\caption{SNe per 0.025 {\it vs.}~$\Sigma \equiv [\sigma^2(B^{\rm raw}_{\rm max}) + 4\sigma^2(V^{\rm raw}_{\rm max}) + \sigma^2(B^{\rm raw}_{BV})]^{1/2}$.  The arrow shows the position of the cut $\Sigma < 0.25$.} \label{sigdB}
\end{figure}


For completeness, Table \ref{cut} lists a final cut that will be justified in \S \ref{global}.  Unlike all cuts discussed so far, the final cut takes into account the residuals from fits that demand consistency between maximum-luminosity color $E$ and {\sc cmagic} color ${\cal E}$ (\S \ref{DeltaE}), and from Hubble fits to the average of $B_{\rm max}$ and $B_{BV}$ (\S \ref{Bavg}).  It excludes as outliers five of the 61 SNe in the plotted sample, yielding a ``fitted sample'' of 56 SNe.  These outliers are excluded from the $\Delta E = 0$ and Hubble fits, but they continue to be included in the figures, identified by open-circle symbols.  


\tabletypesize{\footnotesize}
\setlength{\tabcolsep}{1.3pt}
\begin{deluxetable*}{lcclllllccc}
\tablecolumns{11}
\tableheadfrac{}
\tablecaption{Results of light-curve and {\sc cmagic} fits.\label{table1a}}
\tabletypesize{\scriptsize}
\tablehead{\colhead{SN}
&\colhead{\begin{tabular}{c} E or \\ S0 \\ host \end{tabular}}
&\colhead{\begin{tabular}{c} No.~of \\ params \\ in {\sc lc} fit \end{tabular}}
&\colhead{\begin{tabular}{c} $B^{\rm raw}_{\rm max}$ \\ (mag) \end{tabular}}
&\colhead{\begin{tabular}{c} $V^{\rm raw}_{\rm max}$ \\ (mag) \end{tabular}}
&\colhead{\begin{tabular}{c} $\Delta m_{15}$ \\ (mag) \end{tabular}}
&\colhead{\begin{tabular}{c} $B^{\rm raw}_{BV}$ \\ (mag) \end{tabular}}
&\colhead{$\beta_{BV}$}
&\colhead{$z_{\rm cmb}$}
&\colhead{\begin{tabular}{c} E$^{\rm MW}$ \\ $(B\!-\!V)$ \\ (mag) \end{tabular}}
&\colhead{Ref}
}
\startdata
1990N&$$&$4$&$12.537(031)$&$12.626(011)$&$1.134(023)$&$13.831(026)$&$2.123(113)$&$0.00447$&$0.026$&$(1)$\\
1990O&$$&$4$&$16.223(056)$&$16.248(035)$&$0.937(059)$&$17.575(021)$&$2.118(059)$&$0.03060$&$0.093$&$(2)$\\
1990af&$\star$&$4$&$17.767(018)$&$17.728(012)$&$1.594(030)$&$18.965(052)$&$2.038(190)$&$0.04992$&$0.035$&$(2)$\\
1991T&$$&$6$&$11.412(054)\tablenotemark{a}$&$11.455(009)$&$1.183(084)\tablenotemark{a}$&$12.415(082)$&$2.029(236)$&$0.00693$&$0.022$&$(3)$\\
1991U&$$&$4$&$16.487(102)$&$16.577(053)$&$1.172(044)$&$17.646(029)$&$1.961(079)$&$0.03096$&$0.062$&$(2)$\\
1991ag&$$&$6$&$14.452(060)$&$14.419(073)$&$0.896(033)$&$15.807(050)$&$1.982(116)$&$0.01388$&$0.062$&$(2)$\\
1992A&$\star$&$4$&$12.609(018)$&$12.489(018)$&$1.349(017)$&$13.654(040)$&$2.006(133)$&$0.00594$&$0.017$&$(4)$\\
1992ag&$$&$6$&$16.250(058)$&$16.152(033)$&$1.080(066)$&$17.226(135)$&$1.842(257)$&$0.02734$&$0.097$&$(2)$\\
1992al&$$&$4$&$14.479(017)$&$14.531(014)$&$1.217(017)$&$15.923(012)$&$2.179(032)$&$0.01350$&$0.034$&$(2)$\\
1992bc&$$&$6$&$15.107(039)$&$15.142(019)$&$0.750(024)$&$16.816(035)$&$2.074(116)$&$0.01978$&$0.022$&$(2)$\\
1992bg&$$&$4$&$16.753(046)$&$16.770(026)$&$1.160(039)$&$18.097(031)$&$2.189(092)$&$0.03648$&$0.185$&$(2)$\\
1992bh&$$&$4$&$17.614(031)$&$17.546(015)$&$0.972(078)$&$18.730(033)$&$2.112(092)$&$0.04530$&$0.022$&$(2)$\\
1992bk&$\star$&$4$&$18.259(068)$&$18.214(049)$&$1.631(043)$&$19.377(028)$&$1.720(122)$&$0.05885$&$0.015$&$(2)$\\
1992bl&$$&$4$&$17.409(055)$&$17.374(040)$&$1.502(045)$&$18.661(060)$&$2.276(183)$&$0.04223$&$0.011$&$(2)$\\
1992bo&$\star$&$4$&$15.748(021)$&$15.754(019)$&$1.631(019)$&$16.925(027)$&$1.838(098)$&$0.01811$&$0.027$&$(2)$\\
1992bp&$$&$4$&$18.258(029)$&$18.308(022)$&$1.354(031)$&$19.605(056)$&$1.721(153)$&$0.07858$&$0.069$&$(2)$\\
1993B&$$&$4$&$18.437(062)$&$18.431(038)$&$1.321(047)$&$19.610(055)$&$1.867(194)$&$0.07009$&$0.079$&$(2)$\\
1993H&$$&$4$&$16.709(023)$&$16.528(018)$&$1.566(026)$&$17.403(033)$&$2.039(090)$&$0.02507$&$0.060$&$(2)$\\
1993O&$\star$&$4$&$17.615(025)$&$17.672(021)$&$1.278(027)$&$18.966(035)$&$2.149(187)$&$0.05293$&$0.053$&$(2)$\\
1993ag&$\star$&$4$&$17.803(028)$&$17.722(023)$&$1.283(039)$&$18.861(049)$&$1.983(243)$&$0.05004$&$0.112$&$(2)$\\
1994D&$\star$&$4$&$11.762(019)\tablenotemark{a}$&$11.855(019)\tablenotemark{a}$&$1.555(027)\tablenotemark{a}$&$13.223(048)$&$2.089(125)$&$0.00261$&$0.022$&$(5)$\\
1994M&$\star$&$4$&$16.324(034)$&$16.265(021)$&$1.409(033)$&$17.500(111)$&$1.993(454)$&$0.02431$&$0.024$&$(6)$\\
1994Q&$\star$&$4$&$16.518(108)$&$16.428(060)$&$0.961(061)$&$17.599(041)$&$2.033(200)$&$0.02897$&$0.017$&$(6)$\\
1994S&$$&$4$&$14.808(029)$&$14.807(041)$&$0.869(099)$&$16.208(026)$&$1.989(067)$&$0.01616$&$0.021$&$(6)$\\
1994ae&$$&$4$&$13.117(055)\tablenotemark{a}$&$12.985(032)$&$1.072(052)\tablenotemark{a}$&$14.402(028)$&$2.050(122)$&$0.00539$&$0.031$&$(6)$\\
1995D&$\star$&$4$&$13.230(016)$&$13.264(014)$&$0.977(015)$&$14.610(030)$&$2.178(117)$&$0.00654$&$0.058$&$(6)$\\
1995ac&$$&$6$&$17.056(032)\tablenotemark{a}$&$17.094(032)\tablenotemark{a}$&$0.852(070)\tablenotemark{a}$&$18.444(026)$&$2.050(127)$&$0.04882$&$0.042$&$(6)$\\
1995ak&$$&$4$&$16.056(046)$&$16.056(047)$&$1.486(054)$&$17.184(038)$&$1.929(174)$&$0.02300$&$0.043$&$(6)$\\
1995al&$$&$4$&$13.319(017)$&$13.187(018)$&$0.964(017)$&$14.318(038)$&$2.027(116)$&$0.00587$&$0.014$&$(6)$\\
1995bd&$$&$6$&$15.220(016)$&$14.910(016)$&$0.776(035)$&$16.172(062)$&$2.163(093)$&$0.01599$&$0.495$&$(6)$\\
1996C&$$&$4$&$16.617(077)\tablenotemark{a}$&$16.528(041)\tablenotemark{a}$&$0.940(038)\tablenotemark{a}$&$17.856(040)\tablenotemark{a}$&$2.182(115)\tablenotemark{a}$&$0.03007$&$0.014$&$(6)$\\
1996X&$\star$&$6$&$12.978(018)\tablenotemark{a}$&$13.007(018)\tablenotemark{a}$&$1.307(021)\tablenotemark{a}$&$14.271(017)$&$2.037(042)$&$0.00800$&$0.069$&$(6,7)$\\
1996bl&$$&$4$&$16.677(029)$&$16.627(021)$&$0.851(038)$&$17.938(024)$&$1.928(068)$&$0.03485$&$0.105$&$(6)$\\
1996bo&$$&$4$&$15.838(023)$&$15.515(012)$&$1.171(030)$&$16.346(021)$&$2.096(091)$&$0.01632$&$0.078$&$(6)$\\
1996z&$$&$4$&$14.409(091)$&$13.992(098)$&$1.060(073)$&$14.901(022)$&$2.487(387)$&$0.00796$&$0.063$&$(6)$\\
1997E&$\star$&$4$&$15.108(019)$&$15.055(013)$&$1.394(028)$&$16.233(014)$&$1.836(062)$&$0.01332$&$0.124$&$(8)$\\
1997bp&$$&$4$&$13.891(026)\tablenotemark{a}$&$13.761(026)\tablenotemark{a}$&$1.288(109)\tablenotemark{a}$&$14.651(065)\tablenotemark{a}$&$1.831(121)\tablenotemark{a}$&$0.00944$&$0.044$&$(8)$\\
1997bq&$$&$4$&$14.442(025)$&$14.342(024)$&$1.160(020)$&$15.276(018)$&$1.994(049)$&$0.00960$&$0.024$&$(8)$\\
1997br&$$&$6$&$13.596(028)\tablenotemark{a}$&$13.418(028)\tablenotemark{a}$&$1.139(034)\tablenotemark{a}$&$14.026(072)$&$2.313(148)$&$0.00630$&$0.113$&$(9)$\\
1997cw&$$&$6$&$15.863(067)$&$15.586(081)$&$0.997(034)$&$16.488(030)$&$1.823(083)$&$0.01595$&$0.073$&$(8)$\\
1998ab&$$&$6$&$16.137(076)\tablenotemark{a}$&$16.113(076)\tablenotemark{a}$&$1.098(035)\tablenotemark{a}$&$17.200(024)$&$2.247(106)$&$0.02784$&$0.017$&$(8)$\\
1998bu&$$&$6$&$12.130(015)$&$11.774(010)$&$0.977(024)$&$12.842(039)$&$2.015(088)$&$0.00416$&$0.025$&$(10)$\\
1998dm&$$&$4$&$14.761(042)$&$14.551(029)$&$0.986(032)$&$15.392(036)$&$2.146(115)$&$0.00556$&$0.044$&$(8)$\\
1998ec&$$&$4$&$16.309(059)$&$16.134(056)$&$1.062(031)$&$17.088(027)$&$2.051(079)$&$0.02012$&$0.085$&$(8)$\\
1998es&$$&$6$&$13.836(013)$&$13.760(008)$&$0.742(015)$&$15.099(031)$&$1.654(177)$&$0.00957$&$0.032$&$(8)$\\
1999aa&$$&$6$&$14.741(014)$&$14.722(009)$&$0.810(019)$&$16.165(008)$&$1.801(029)$&$0.01525$&$0.040$&$(11)$\\
1999aw&$$&$6$&$16.694(028)\tablenotemark{a}$&$16.745(019)\tablenotemark{a}$&$0.787(026)\tablenotemark{a}$&$18.245(015)$&$1.736(071)$&$0.03924$&$0.032$&$(12)$\\
1999dk&$$&$4$&$14.817(027)$&$14.783(014)$&$1.002(029)$&$15.880(025)$&$2.148(076)$&$0.01395$&$0.054$&$(13)$\\
1999dq&$$&$6$&$14.405(016)$&$14.342(009)$&$0.953(028)$&$15.475(055)$&$2.074(106)$&$0.01308$&$0.110$&$(8)$\\
1999ee&$$&$6$&$14.853(007)$&$14.566(007)$&$0.904(008)$&$15.650(014)$&$2.081(032)$&$0.01055$&$0.020$&$(14)$\\
2000E&$$&$6$&$12.784(064)\tablenotemark{a}$&$12.757(064)\tablenotemark{a}$&$1.070(115)\tablenotemark{a}$&$13.884(010)$&$1.752(015)$&$0.00422$&$0.364$&$(15)$\\
2000ce&$$&$4$&$16.961(061)$&$16.518(030)$&$0.961(032)$&$17.152(032)$&$2.183(059)$&$0.01650$&$0.057$&$(13)$\\
2000cf&$$&$4$&$17.087(059)$&$17.103(018)$&$1.316(056)$&$18.327(034)$&$2.068(111)$&$0.03646$&$0.032$&$(8)$\\
2000cn&$$&$4$&$16.601(009)$&$16.399(010)$&$1.582(027)$&$17.322(036)$&$1.972(389)$&$0.02321$&$0.057$&$(8)$\\
2000dk&$\star$&$4$&$15.354(027)\tablenotemark{a}$&$15.330(012)$&$1.440(042)\tablenotemark{a}$&$16.424(030)$&$1.971(157)$&$0.01645$&$0.070$&$(8)$\\
2001V&$$&$6$&$14.606(064)\tablenotemark{a}$&$14.546(064)$&$0.730(081)\tablenotemark{a}$&$15.766(086)$&$1.858(286)$&$0.01604$&$0.020$&$(16)$\\
2001ba&$$&$4$&$16.322(025)$&$16.342(014)$&$0.911(024)$&$17.761(059)$&$2.367(255)$&$0.03053$&$0.064$&$(17)$\\
2001el&$$&$4$&$12.793(027)\tablenotemark{a}$&$12.691(027)\tablenotemark{a}$&$1.161(022)\tablenotemark{a}$&$13.621(008)$&$2.092(028)$&$0.00365$&$0.014$&$(18)$\\
2002bo&$$&$4$&$13.927(018)$&$13.591(016)$&$1.335(032)$&$14.395(067)$&$1.874(127)$&$0.00547$&$0.025$&$(19)$\\
2002el&$$&$4$&$16.153(025)$&$16.179(020)$&$1.375(025)$&$17.381(044)$&$1.961(139)$&$0.02238$&$0.085$&$(20)$\\
2002er&$$&$4$&$14.266(012)$&$14.108(012)$&$1.253(010)$&$15.070(037)$&$1.984(079)$&$0.00855$&$0.157$&$(21)$\\
\enddata
\tablenotetext{a}{The uncertainty in this quantity has been adjusted as discussed in Appendix \ref{appLC}.}
\tablerefs{(1) \citet{Lira:1998}; (2) \citet{Hamuy:1996b}; (3) \citet{Altavilla:2004}; (4) \citet{Suntzeff:1996}; (5) \citet{Richmond:1995}; (6) \citet{Riess:1999}; (7) \citet{Salvo:2001}; (8) \citet{Jha:2002}; (9) \citet{Li:1999}; (10) \citet{Jha:1999}; (11) \citet{Krisciunas:2000}; (12) \citet{Strolger:2002}; (13) \citet{Krisciunas:2001}; (14) \citet{Stritzinger:2002}; (15) \citet{Valentini:2003}; (16) \citet{Vinko:2003}; (17) \citet{Krisciunas:2004}; (18) \citet{Krisciunas:2003}; (19) \citet{Benetti:2004}; (20) Wang, L.\ \& Li, W.D.\ 2005, in preparation; (21) \citet{Pignata:2004}.}
\end{deluxetable*}


Table \ref{table1a} presents the results of our {\sc lc} and {\sc cmagic} fits to the plotted sample.  Tabulated are the SN name; the symbol ``$\star$'' if the host galaxy is E, E/S0, or S0; the number of adjustable parameters in the {\sc lc} fit to one band (see \S \ref{LCfits}); the raw peak luminosities in the $B$ and $V$ bands; the decline rate $\Delta m_{15}$; the raw {\sc cmagic} luminosity $B^{\rm raw}_{BV}$ and slope $\beta_{BV}$; the redshift $z$ in the {\sc cmb} frame; the excess $\bv$ color ${\rm E}^{\rm MW}(B \!-\! V)$ due to Milky Way dust in the line of sight to the SN; and the photometry references that supplied the data for our fits.  All fit quantities have been $K$-corrected.  The tabulated uncertainties were propagated from the published photometry errors; uncertainties that are flagged in Table \ref{table1a}'s footnote were adjusted as discussed in Appendix \ref{appLC}.  The origins of the SNe in Table \ref{table1a} are diverse, but they are drawn mainly from \citet{Hamuy:1996b}; \citet{Riess:1999}; and \citet{Jha:2002}.  The SN fluxes that were input to our {\sc lc} and {\sc cmagic} fits were taken from the compilation used by Conl05.

All $B^{\rm raw}_{\rm max}$,  $V^{\rm raw}_{\rm max}$, and $B^{\rm raw}_{BV}$ magnitudes in Table \ref{table1a} have been corrected for {\sc mw} dust extinction using the dust map of \citet{Schlegel:1998} and the law of Card89.  The corrections to $B^{\rm raw}_{\rm max}$ and $V^{\rm raw}_{\rm max}$ used extinction coefficients $R_B = 4.15$ and $R_V = 3.14$, respectively, while, following equation (\ref{eqA}), the correction to $B^{\rm raw}_{BV}$ used a coefficient that was smaller than $R_B$ by the value of $\beta_{BV}$ measured for the particular SN being corrected.  More generally, quantities appearing anywhere in this paper that require correction for the effects of {\sc mw} dust, whether they are numbers or symbols in equations, should be assumed already to have been corrected for these effects.  If necessary, our numerical corrections can be undone by using the values of ${\rm E}^{\rm MW}(B \!-\! V)$ found in Table \ref{table1a}.

To help visualize the {\sc lc} fit results, Fig.~\ref{ridge} displays a scatter plot of maximum-luminosity color $E$ {\it vs.}~$\Delta m_{15} \, $.  As detailed in the caption, different symbols divide the sample by host-galaxy morphology.  In \S \ref{blue} the characteristics of the portion of Fig.~\ref{ridge} in which SNe are clustered at $|E| \la 0.1$ are explored.  


\begin{figure}
\epsscale{1.17}
\plotone{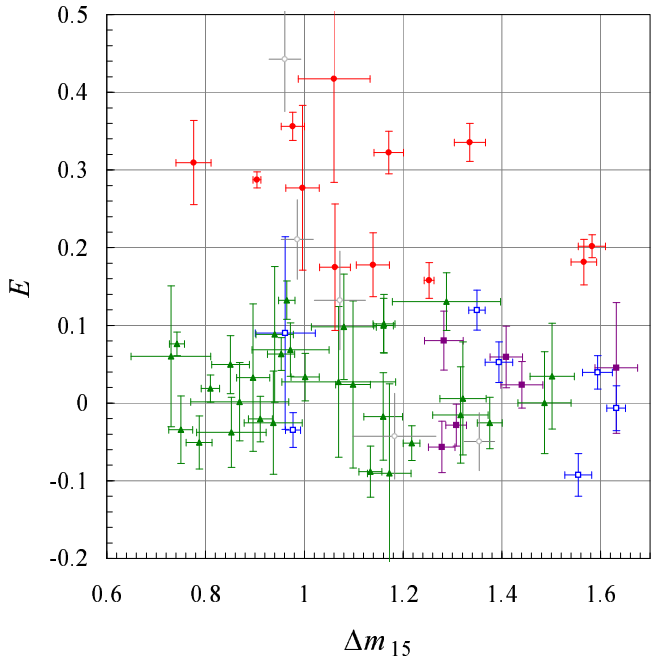}
\caption{Maximum-luminosity color $E \equiv B^{\rm raw}_{\rm max} - V^{\rm raw}_{\rm max}$ {\it vs.}~decline rate $\Delta m_{15}$ calculated from our light curve fits.  Closed squares (colored violet in electronic edition) show SNe from elliptical (E) and E/S0 host galaxies; open squares (colored blue in electronic edition) show SNe from S0 hosts.  SNe from remaining (late-type) hosts are divided into those with normal color, shown by closed triangles (colored green in electronic edition); and those that are reddened, shown by closed circles (colored red in electronic edition).  Open circles depict SNe that do not participate in the global fit.} \label{ridge}
\end{figure}


\section{COMPARISONS OF $B_{\rm max}$ TO $B_{BV}$} \label{DeltaE} 

$B_{\rm max}$ can be compared to the {\sc cmagic} output $B_{BV}$ with substantially higher precision than is achieved by subjecting either of these luminosity measures to its own Hubble fit.  Primarily this is true because $B_{\rm max} - B_{BV}$ is immune to fluctuations in SN peculiar velocity that, in our range $z < 0.1$, are a major source of uncertainty in Hubble fits.  (As well, $B_{\rm max} - B_{BV}$ is blind to cosmological variations in Hubble's law.)  After correction for observed SN color has been performed (as is implied thoughout this paper if the $^{\rm raw}$ notation is absent), $B_{\rm max} - B_{BV}$ has the additional advantage that it depends only multiplicatively upon the effective color-correction coefficient ${\cal R}$, a phenomenological parameter that must separately be fit.  As eqs.~(\ref{eqC}) and (\ref{eqE}) show, one may divide $B_{\rm max} - B_{BV}$ by ${\cal R}$ to obtain the difference $\Delta E \equiv {\cal E} - E$ between the {\sc cmagic} color ${\cal E} \equiv (B^{\rm raw}_{\rm max} - B^{\rm raw}_{BV})/\beta_{BV}$ and the maximum-luminosity color $E \equiv B^{\rm raw}_{\rm max}-V^{\rm raw}_{\rm max}$.  It is to the study of $\Delta E$, which is completely independent of ${\cal R}$ as well as SN peculiar velocity, that this section is devoted.  A drawback of studying $\Delta E$ is that any systematic variation it may exhibit cannot, without more information, be apportioned uniquely between $B_{\rm max}$ and $B_{BV}$.  

Because the observed SN color is influenced by intrinsic SN color as well as intervening dust, and because $B_{\rm max}$ and $B_{BV}$ probe epochs at which intrinsic SN color might possibly be manifested differently, in principle $\Delta E$ might vary systematically with observed color itself.  After correcting $\Delta E$ for systematic dependence on SN decline rate, to which we shall return shortly, we tested this possibility by fitting $\Delta E$ to a linear function of $E$.  The additional free parameter allowed $\chi^2$ to diminish by only 0.1, lending no support to this possibility.  Accordingly, in the rest of this paper, we make no systematic color correction to $\Delta E$ itself, and, in results that are based on eqs.~(\ref{eqC}) and (\ref{eqE}), we draw no distinction between values of ${\cal R}$ used for calculating corrected $B_{\rm max}$ as opposed to $B_{BV}$.

\subsection{Decline-Rate Dependence of Color Difference} \label{declineDeltaE}

Figure \ref{dogleg} shows the dependence upon the decline-rate parameter $\Delta m_{15}$ of the color difference $\Delta E$.  Though we were aware of the gentle nonlinearity in $\Delta m_{15}$ of Hubble residuals for $B$ and $V$ already reported by Phil99, we were unprepared for the striking appearance of Fig.~\ref{dogleg}, which has acquired the nickname ``dogleg plot''.  The data there are fit to a bilinear parametrization 
\begin{equation}
0 = \Delta E -\alpha_{\Delta E}(\Delta m_{15} - 1.1) -
  \alpha'_{\Delta E} |\Delta m_{15}-1.1| + {\rm const} \; , \label{eqF}
\end{equation}
with a ``kink'' at the nominal centroid $\Delta m_{15} = 1.1$ {mag}.  (Evidently, for fast decliners ($\Delta m_{15} > 1.1$) the decline-rate coefficient is $\alpha_{\Delta E} + \alpha'_{\Delta E}$, and for slow decliners it is $\alpha_{\Delta E} - \alpha'_{\Delta E}$.)  If only photometric errors on $\Delta E$ are taken into account, $\chi^2$ is unacceptably large (99.5 for 53 degrees of freedom (dof)).  Adding in quadrature to the photometric errors a modest rms noise ({\it i.e.}~intrinsic scatter) term $n_{\Delta E} = 0.029$ {mag} reduces $\chi^2$ to its most probable value of 51, where its probability density is larger by the factor $\exp{(7.2)}$.  The same likelihood ratio would be associated with a gaussian fluctuation of $\sqrt{14.4} = 3.8$ standard deviations (3.8$\sigma$).  Therefore, in the difference between {\sc cmagic} and maximum-luminosity ${\bv}$ color, intrinsic noise clearly is present at a modest level ($0.029 \pm 0.007$ {mag}).


\begin{figure}
\epsscale{1.17}
\plotone{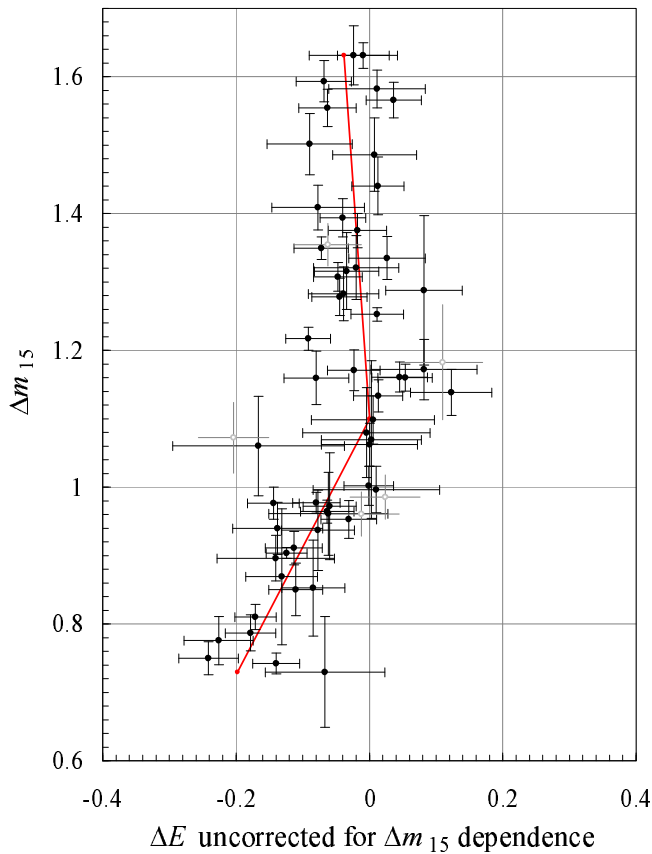}
\caption{Decline rate $\Delta m_{15}$ ({mag}) {\it vs.}~the difference $\Delta E$ ({mag}) between {\sc cmagic} and maximum-luminosity colors.  In this plot the abscissa is uncorrected for decline-rate dependence.  The fit line is the function in equation (\ref{eqF}).  SNe marked by open circles are omitted from the fit.}\label{dogleg}
\end{figure}


Only after noise is added to bring $\chi^2$ down to a probable level ($\chi^2/{\rm dof} \to 1$ as ${\rm dof} \to \infty$) can the errors on the fit parameters be trusted.  These parameters and their errors are shown in the upper portion of Table \ref{Dm15}.  Various alternate conditions for the fit are exhibited in the lower portion of the same table, together with the corresponding $\Delta \chi^2$ penalty that is assessed when each condition is enforced.  The $\Delta E$ data definitely reject the hypothesis of no $\Delta m_{15}$ dependence at all ($\Delta \chi^2 = 86.1$ for 2 dof).  Fitting the data to the bilinear parametrization from equation (\ref{eqF}), both a nonzero linear coefficient $\alpha_{\Delta E}$ (7.5$\sigma$) and a nonzero kink coefficient $\alpha'_{\Delta E}$ (5.8$\sigma$) are required.   Therefore, in $\Delta E$ the presence of a nonlinear decline-rate dependence is firmly established.


\tabletypesize{\footnotesize}
\setlength{\tabcolsep}{5pt}
\begin{deluxetable}{lr}
\tablewidth{0pt}
\tablecolumns{2}
\tableheadfrac{}
\tablecaption{Results of bilinear fit to data in Fig.~\ref{dogleg}.\label{Dm15}}
\tabletypesize{}
\tablehead{\colhead{\begin{tabular}{l} Result of $\Delta E$ fit \end{tabular}}
&\colhead{\begin{tabular}{r} Fit value \end{tabular}}
}
\startdata
$n_{\Delta E}$&$0.029 \pm 0.007$\\
$\alpha_{\Delta E}$&$0.23 \pm 0.03$\\
$\alpha'_{\Delta E}$&$-0.30 \pm 0.05$\\
$\chi^2$ / dof&51.0 / 53\\
\cutinhead{Fit conditions\tablenotemark{a} $\;\;\;\;\;\;\;\;\;\;\;\;\;\;\;\;\;\;$ $\Delta \chi^2$}
default (as above)&0\\
$n_{\Delta E} = 0$&+48.5\\
$\alpha_{\Delta E} = 0$&+57.1\\
$\alpha'_{\Delta E} = 0$&+33.6\\
$\alpha_{\Delta E} = \alpha'_{\Delta E} = 0$&+86.1\\
quadratic (not bilinear)&+7.5\\
\enddata
\tablenotetext{a}{Other parameters are allowed to vary.}
\end{deluxetable}


Especially because the bilinear parametrization is nonstandard, it is natural to explore alternatives.  Obviously, the first derivative with respect to $\Delta m_{15}$ of any parametrization ideally should be continuous.  For the bilinear case this could be accomplished by substituting a small curved segment for the corner at $\Delta m_{15} = 1.1$.  Here we omit this feature because the data near the corner are too sparse to determine the curvature there.  The most common nonlinear form is a quadratic.  As recorded in the final row of Table \ref{Dm15}, a $\Delta \chi^2$ penalty of $+7.5$ is paid for substituting a quadratic function of $\Delta m_{15}$ for the bilinear form.  Statistically, with respect to the bilinear form, this sets the betting odds against the quadratic at 43:1.  Another factor weighing against a quadratic parametrization is its asymptotic behavior: the data in Fig.~\ref{dogleg} do not appear to prefer a steeper slope in $\Delta m_{15}$ as the extremes in that parameter are approached.  For all further analysis in this paper we continue to parametrize the $\Delta m_{15}$ dependence by a bilinear function like that in equation (\ref{eqF}).

With intrinsic noise and nonlinear decline-rate dependence established as basic features of $B_{\rm max}$~$-$~$B_{BV}$ by the results in Fig.~\ref{dogleg} and Table \ref{Dm15}, the key issue not yet addressed is raised: how is responsibility for these features apportioned between $B_{\rm max}$ and $B_{BV}$?  This statistically challenging issue must be attacked using Hubble fits, to which we now turn.

\section{HUBBLE FIT TO AVERAGE OF $B_{\rm max}$ AND $B_{BV}$} \label{Bavg}

With a relatively precise fit to the difference of $B_{\rm max}$ and $B_{BV}$ now in hand, we choose as the input to our primary Hubble fit the orthogonal linear combination of these two magnitudes, {\it i.e.}~their unweighted average $\langle B \rangle$.  As will be demonstrated in \S \ref{global}, the sum of $\chi^2$ for both of these fits is equivalent to the $\chi^2$ for a global fit to all of the $B_{\rm max}$ and $B_{BV}$ data.  Although this is the simplest and most straightforward type of global fit, it is not the only possible approach:  for example, one could set aside the fit to $B_{\rm max} - B_{BV}$ and substitute simultaneous Hubble fits to $B_{\rm max}$ and $B_{BV}$.  That approach would have required taking careful account of the $\{B_{\rm max},B_{BV}\}$ covariance matrix for each SN, including the large off-diagonal term due to peculiar velocity.  Using that approach, the striking decline-rate dependence shown in Fig.~\ref{dogleg} could have been reproduced, but the data supporting it would have been buried in the details of those matrices.

All Hubble fits performed in this paper assume a flat universe with the ratio $\Omega_M$ of mass to critical densities set at 0.3.  Over the range in redshift of the SNe we consider, $z < 0.1$, the results of these fits are insensitive to variations in $\Omega_M$ and $\Omega_\Lambda$ that reasonably would be allowed by current world data.  Also, for every Hubble fit that is reported here, as well as for every fit to $\Delta E = 0$, a constant overall magnitude offset is included that always is allowed to assume its own best fit value, independent of the offset determined by any other fit.  (For Hubble fits, this offset is essentially the absolute SN magnitude combined with the Hubble constant.)  The best fit values of these offsets are left unreported in this paper since they are irrelevant to its goals.  Because only linear color corrections are used here, any constant offset in color plays the same role as a constant overall offset and therefore leaves unaffected any fit results that we report.  Accordingly this paper does not distinguish between a color $\bv$ and its excess with respect to an expected color.
 
Apart from the assumptions implicit in using particular $B$ and $V$ light-curve templates, and in exploiting the observed linear relationship between $B$ magnitude and $\bv$ color in the {\sc cmagic} region, all analysis in this paper makes no prior assumption.  Specifically, because the effects of extinction by host-galactic dust and of intrinsic SN color variation are not yet well characterized, we accept and use the observed $\bv$ color without applying to it any correction based {\it e.g.}~on a Bayesian prior.  The statistical error analysis applied here is frequentist:  in a multiparameter fit, when a single parameter is varied and the change in likelihood is recorded, all other parameters (including the offset) are allowed also to vary for best compatibility with that excursion.  (If instead we had chosen to integrate over a likelihood surface, implicitly we would have had to apply a Bayesian prior.)  Thus our calculated errors are never assumed to be symmetric; when an average error is reported, the calculated errors lie within $\pm 5\%$ of it.
 
\subsection{Correcting ${\langle B \rangle}$ for Color and Decline Rate} \label{BavgCorr}

Because $\Delta B \equiv B_{\rm max} - B_{BV}$ is proportional to $\Delta E$ (eqs.~(\ref{eqC}) and (\ref{eqE})), the bilinear dependence of $\Delta E$ upon $\Delta m_{15}$ should map into a bilinear $\Delta m_{15}$-dependent correction to $B_{\rm max}$, $B_{BV}$, or both.  The Hubble fit described here allows for both.  Allowing for corrections for both color and decline-rate dependence, the more general form of eqs.~(\ref{eqC}) and (\ref{eqE}) becomes
\begin{eqnarray}
B_{\rm max} &=& B^{\rm raw}_{\rm max} - {\cal R}E 
  - {\cal D}(\Delta m_{15}; \alpha_{\rm max},\alpha'_{\rm max}) \label{eqG} \\
B_{BV} &=& B^{\rm raw}_{\rm max} - {\cal R}{\cal E} 
  - {\cal D}(\Delta m_{15}; \alpha_{BV},\alpha'_{BV}) \; , \label{eqH}
\end{eqnarray}
where the function ${\cal D}(\Delta m_{15}; \alpha,\alpha')$ is shorthand for the bilinear decline-rate dependence introduced in equation (\ref{eqF}): 
\begin{equation}
{\cal D}(\Delta m_{15}; \alpha,\alpha') \equiv
  \alpha(\Delta m_{15} - 1.1) + \alpha'|\Delta m_{15}-1.1| \; . \label{eqHI}
\end{equation}   
Apart from the ``kink'' term in $\alpha'_{\rm max}$, the corrections described by equation (\ref{eqG}) are equivalent to those introduced by \citet{Tripp:1998}.  The coefficients $\alpha_{BV}$ and $\alpha'_{BV}$ for $B_{BV}$ are not the same as the coefficients $\alpha_{\rm max}$ and $\alpha'_{\rm max}$ for $B_{\rm max}$; rather, the two pairs of coefficients are linked by the decline-rate corrections to $\Delta E \equiv {\cal E} - E$ given by equation (\ref{eqF}).  Straightforwardly, the average ${\langle B \rangle}$ of $B_{\rm max}$ and $B_{BV}$ is corrected for color and decline rate by a similar relation:
\begin{equation}
\langle B \rangle = B^{\rm raw}_{\rm max} - {\cal R}\langle E \rangle 
  - {\cal D}(\Delta m_{15}; 
  \alpha_{\langle B \rangle},\alpha'_{\langle B \rangle}) \; , \label{eqI} \\
\end{equation}
where $\langle E \rangle \equiv (E + {\cal E})/2$ and the decline-rate coefficients $\alpha_{\langle B \rangle}$  and $\alpha'_{\langle B \rangle}$ are equal to the average of those for $B_{\rm max}$ and $B_{BV}$.  Table \ref{Dm152} provides a full list of the bilinear decline-rate coefficients and their interrelationships.  There, in terms of the color coefficient ${\cal R}$ and of the decline-rate coefficients for correcting $\Delta E$ and $B_{\rm max}$, the last column supplies the decline-rate coefficients used to correct $\Delta B$, $\langle B \rangle$, and $B_{BV}$.


\tabletypesize{\footnotesize}
\setlength{\tabcolsep}{5pt}
\begin{deluxetable}{lccr}
\tablewidth{0pt}
\tablecolumns{4}
\tableheadfrac{}
\tablecaption{Definitions of bilinear decline-rate correction coefficients used in the fit to $\Delta E = 0$; used for correcting $\Delta B$; and used in Hubble fits to $B_{\rm max} = \mu$, $\langle B \rangle = \mu$, and $B_{BV} = \mu$.\label{Dm152}}
\tabletypesize{}
\tablehead{\colhead{\begin{tabular}{l} Coef- \\ ficient \end{tabular}}
&\colhead{\begin{tabular}{c} Multiplying \end{tabular}}
&\colhead{\begin{tabular}{c} To \\ correct \end{tabular}}
&\colhead{\begin{tabular}{c} Is equivalent \\ to \end{tabular}}
}
\startdata
$\alpha_{\Delta E}$&$(\Delta m_{15}  - 1.1)$&$\Delta E$&\nodata\\
$\alpha'_{\Delta E}$&$|\Delta m_{15}  - 1.1|$&$\Delta E$&\nodata\\
$\alpha_{\Delta B}$&$(\Delta m_{15}  - 1.1)$&$\Delta B$&${\cal R}\alpha_{\Delta E}$\\
$\alpha'_{\Delta B}$&$|\Delta m_{15}  - 1.1|$&$\Delta B$&${\cal R}\alpha'_{\Delta E}$\\
$\alpha_{\rm max}$&$(\Delta m_{15}  - 1.1)$&$B_{\rm max}$&\nodata\\
$\alpha'_{\rm max}$&$|\Delta m_{15}  - 1.1|$&$B_{\rm max}$&\nodata\\
$\alpha_{\langle B \rangle}$&$(\Delta m_{15}  - 1.1)$&$\langle B \rangle$&$\alpha_{\rm max} - {{\cal R} \over {2}}\alpha_{\Delta E}$\\
$\alpha'_{\langle B \rangle}$&$|\Delta m_{15}  - 1.1|$&$\langle B \rangle$&$\alpha'_{\rm max} - {{\cal R} \over {2}}\alpha'_{\Delta E}$\\
$\alpha_{BV}$&$(\Delta m_{15}  - 1.1)$&$B_{BV}$&$\alpha_{\rm max} - {\cal R}\alpha_{\Delta E}$\\
$\alpha'_{BV}$&$|\Delta m_{15}  - 1.1|$&$B_{BV}$&$\alpha'_{\rm max} - {\cal R}\alpha'_{\Delta E}$\\
\enddata
\end{deluxetable}


For completeness, this section concludes with a brief discussion of the ``effective color'' $E^{\rm eff}$ that multiplies the color-correction coefficient ${\cal R}$ in eqs.~(\ref{eqH}) and (\ref{eqI}).  This detail will be relevant only for visualizing the quality of the fit that determines ${\cal R}$.  In eqs.~(\ref{eqG}), (\ref{eqH}), and (\ref{eqI}), the coefficients ${\cal R}$ are all the same, but they multiply colors -- $E$, ${\cal E}$, and $\langle E \rangle$ -- that obviously are different.  Less obvious is the fact that the right-hand term in each of eqs.~(\ref{eqH}) and (\ref{eqI}) itself is a linear function of ${\cal R}$, through the dependence of $\alpha^{(\prime)}_{BV}$ and $\alpha^{(\prime)}_{\langle B \rangle}$ upon ${\cal R}$ that is shown in the last four rows of Table \ref{Dm152}.  Using the information tabulated there to reexpress $\alpha^{(\prime)}_{BV}$ and $\alpha^{(\prime)}_{\langle B \rangle}$ in terms of $\alpha^{(\prime)}_{\rm max}$ and $\alpha^{(\prime)}_{\Delta E}$, and collecting into $E^{\rm eff}$ the sum of the coefficients of ${\cal R}$ in equation (\ref{eqH}) and in equation (\ref{eqI}), one obtains the effective colors
\begin{mathletters}
\begin{eqnarray}
B_{BV}: \;\;\; E^{\rm eff}_{BV} &=& {\cal E}  
  - {\cal D}(\Delta m_{15}; 
  \alpha_{\Delta E},\alpha'_{\Delta E}) \label{eqK} \\
\langle B \rangle: \;\;\; E^{\rm eff}_{\langle B \rangle} &=& 
  {\langle E \rangle} 
  - {\cal D}(\Delta m_{15}; 
  {\textstyle{\frac{1}{2}}} \alpha_{\Delta E},
  {\textstyle{\frac{1}{2}}} \alpha'_{\Delta E}) \; . \label{eqL}
\end{eqnarray}
\end{mathletters}
When it is uncorrected for color, the Hubble residual {\it e.g.}~of $\langle B \rangle$ should exhibit a linear dependence on $E^{\rm eff}_{\langle B \rangle}$ with slope ${\cal R}$. 

\subsection{Propagating Intrinsic Noise into ${\langle B \rangle}$} \label{noiseProp}

Leaving aside sources of noise that are common to both $B_{\rm max}$ and $B_{BV}$, and introducing a mild assumption that is discussed fully in Appendix \ref{appN}, the rms intrinsic noise $n_{\Delta B}$ that is measured in the comparison of $B_{\rm max}$ to $B_{BV}$ propagates into $\langle B \rangle$ as $n_{\langle B \rangle} = n_{\Delta B}/2$.  Therefore, using the factor ${\cal R}$ to relate the intrinsic noise in $\Delta B$ to that in $\Delta E$, the added rms intrinsic noise $n_{\Delta E} = 0.029 \pm 0.007$ {mag} that was required by the fit to $\Delta E = 0$ (Table \ref{Dm15}) propagates into an added rms intrinsic noise term $n_{\langle B \rangle} = {\cal R} \, n_{\Delta E}/2$ in the average ${\langle B \rangle}$ of $B_{\rm max}$ and $B_{BV}$.

As is also discussed in Appendix \ref{appN}, intrinsic noise that is common to both $B_{\rm max}$ and $B_{BV}$ may be present, together with common-mode noise from the effects of peculiar velocity along the line of sight.  In principle, measuring the redshift dependence of the common-mode noise would allow the effects of peculiar velocity to be resolved from those of other sources.  However, the SN sample statistics available to us do not permit a clean separation.  Instead we resort to parametrizing the sum of all sources of common-mode noise as the manifestation of a single effective rms SN velocity $v_{\rm eff}$ relative to the Hubble flow.  In \S \ref{peculiar} we report the value of $v_{\rm eff}$ determined from the data, from which we deduce an upper limit on the true rms peculiar velocity.    


\begin{figure}
\epsscale{1.17}
\plotone{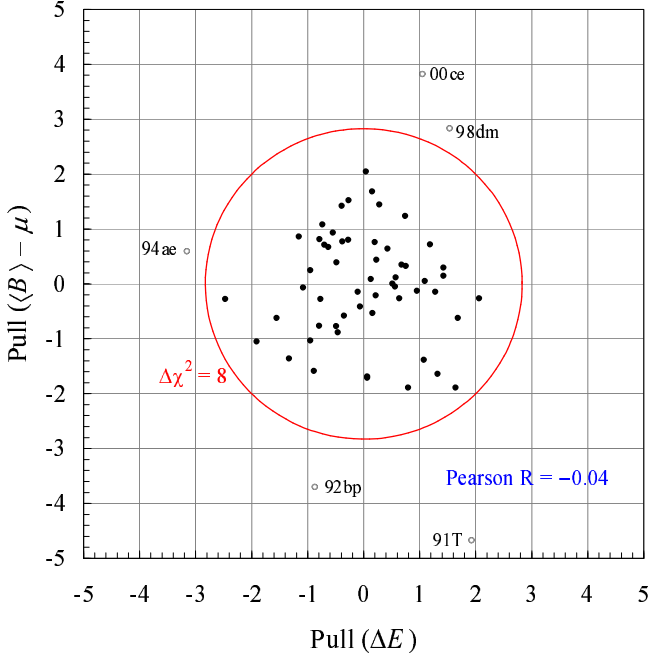}
\caption{Pull ($\equiv$ residual/error) from Hubble fit to $\langle B \rangle$ {\it vs.}~pull from fit to $\Delta E = 0$.  SNe that lie within the circle (filled points) contribute $\Delta \chi^2 < 8$ to the global $\chi^2$ and are included in the fits.} \label{circle}
\end{figure}


\subsection{Equivalent Global Fit and Fitted Sample} \label{global}

Figure \ref{circle} shows for each SN the pull (defined as the residual divided by the error on that residual) for the Hubble fit to ${\langle B \rangle}$ {\it vs.}~the pull for the fit to $\Delta E = 0$ described in \S \ref{declineDeltaE}.  The insignificant correlation between the two pulls, measured by the Pearson correlation coefficient $|{\rm R}| <  0.05$, confirms that, as expected, the two fits are orthogonal.  Therefore the sum of $\chi^2$ for both fits is equivalent to the $\chi^2$ for a global fit to the parameters $\alpha_{\Delta E}$, $\alpha'_{\Delta E}$, $\alpha_{\langle B \rangle}$,  $\alpha'_{\langle B \rangle}$, and ${\cal R}$ (plus offsets).  Note that the decline rate parameters for $B_{\rm max}$ and $B_{BV}$ are also fully specified by the global fit (Table \ref{Dm152}).  One SN contributes two degrees of freedom to the global fit:  one dof for the Hubble fit to ${\langle B \rangle}$ and one dof for the fit to $\Delta E = 0$.  Its contribution to the global $\chi^2$ is equal to the sum of the squares of the same pulls exhibited for that SN in Fig.~\ref{circle}.

If it lies outside the circle of radius $\sqrt{8}$ displayed in Fig.~\ref{circle}, one SN contributes  to the global $\chi^2$ an increment $\Delta \chi^2 > 8$ for 2 dof.  If the pulls were distributed according to gaussian statistics, for our sample size an average of one SN would be expected to lie outside this circle.  In fact five SNe lie outside: four mainly because of a large $|$pull$|$ in the Hubble fit, and one primarily because of a large $|$pull$|$ in the fit to $\Delta E = 0$.  As previously noted in \S \ref{fitPhoto}, we exclude from fits all five of these SNe, leaving a fitted sample of 56 SNe.  Using open-circle symbols we continue to exhibit these outliers in plots.

\subsection{Decline-Rate Correction to Hubble Fit for ${\langle B \rangle}$} \label{declineHub}

In Fig.~\ref{dogleg2} the decline rate $\Delta m_{15}$ is plotted {\it vs.}~the Hubble residual for $\langle B \rangle$.  In this plot $\langle B \rangle$ is fully corrected for the effects of color, as will be discussed in \S \ref{colorHub}, but it is left uncorrected for decline rate.  The line is the correction determined by the global fit and described by equation (\ref{eqI}).  The best-fit kink at $\Delta m_{15} = 1.1$ is less pronounced than that observed for $\Delta E$ (Fig.~\ref{dogleg}).  Because the decline-rate dependence is much smoother for the sum of $B_{\rm max}$ and $B_{BV}$ than for their difference, the kink in decline-rate dependence should (and does) appear with opposite sign in the Hubble residuals for $B_{\rm max}$ and $B_{BV}$; this point will be discussed further in \S \ref{declineBmax}.   In Table \ref{Dm153}, the top two rows list the globally fit coefficients of $\Delta m_{15}$ and $|\Delta m_{15}|$ that correct the Hubble residuals of ${\langle B \rangle}$ via equation (\ref{eqI}); these parameters determine the line in Fig.~\ref{dogleg2}.  

\begin{figure}
\epsscale{1.17}
\plotone{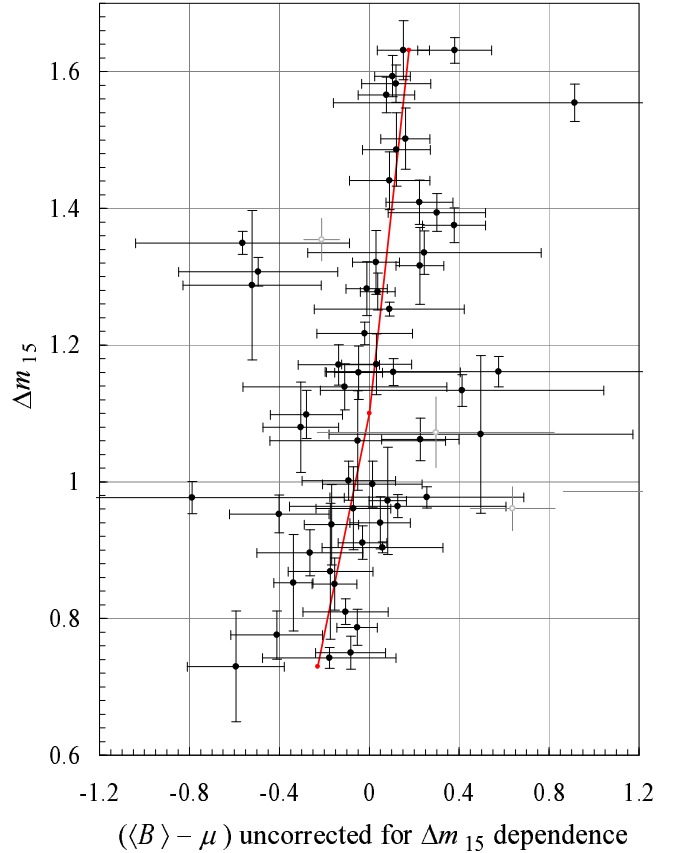}
\caption{Decline rate $\Delta m_{15}$ {\it vs.}~Hubble residual $(\langle B \rangle - \mu)$, which here is uncorrected for its dependence on $\Delta m_{15}$.  The line is a bilinear function of $\Delta m_{15}$ from the global fit, from which SNe marked by open circles are omitted.}\label{dogleg2}
\end{figure}


\tabletypesize{\footnotesize}
\setlength{\tabcolsep}{5pt}
\begin{deluxetable}{lr}
\tablewidth{0pt}
\tablecolumns{2}
\tableheadfrac{}
\tablecaption{Bilinear decline-rate correction coefficients for $\langle B \rangle$, $B_{\rm max}$, and $B_{BV}$ from the global fit.\label{Dm153}}
\tabletypesize{}
\tablehead{\colhead{\begin{tabular}{l} Result of global fit \end{tabular}}
&\colhead{\begin{tabular}{r} Fit value \end{tabular}}
}
\startdata
$\alpha_{\langle B \rangle}$&$  0.48 \pm 0.09$\\
$\alpha'_{\langle B \rangle}$&$-0.14 \pm 0.18$\\
$\alpha_{\rm max}$&$  0.78 \pm 0.09$\\
$\alpha'_{\rm max}$&$-0.54 \pm 0.19$\\
$\alpha_{BV}$&$  0.18 \pm 0.12$\\
$\alpha'_{BV}$&$  0.25 \pm 0.21$\\
$\alpha_{\rm max} - \alpha_{BV}$&$  0.60 \pm 0.09$\\
$\alpha'_{\rm max} - \alpha'_{BV}$&$-0.79 \pm 0.16$\\
\cutinhead{Fit conditions\tablenotemark{a} $\;\;\;\;\;\;\;\;\;\;\;\;\;\;\;\;\;\;$ $\Delta \chi^2$}
default (as above)&0\\
$\alpha'_{\rm max} = 0$&+7.2\\
$\alpha_{\rm max} + \alpha'_{\rm max}= 0$&+2.3\\
$\alpha_{BV} = \alpha'_{BV}= 0$&+9.2\\
$\alpha_{\rm max} - \alpha_{BV}= 0$\tablenotemark{b}&+57.1\\
$\alpha'_{\rm max} - \alpha'_{BV}= 0$\tablenotemark{b}&+33.6\\
\enddata
\tablenotetext{a}{Other parameters are allowed to vary.}
\tablenotetext{b}{From Table \ref{Dm15}.}
\end{deluxetable}


\subsection{Color Correction to Hubble Fit for ${\langle B \rangle}$} \label{colorHub}

Figure \ref{Rcoeff} exhibits the color correction to $\langle B \rangle$.  On the ordinate is plotted the effective color $E^{\rm eff}_{\langle B \rangle}$ from equation (\ref{eqL}); this is the coefficient of ${\cal R}$ in equation (\ref{eqI}).  As in Fig.~\ref{dogleg2}, the horizontal coordinate is the Hubble residual for ${\langle B \rangle}$}; however, in this plot $\langle B \rangle$ is fully corrected for the effects of decline rate, but it is left uncorrected for color.  Again, the line is determined by the global fit; in this case the correction is purely linear, with best fit slope ${\cal R} = 2.59 \pm 0.24$.  This result is inconsistent either with a null correction or with a coefficient typical of Milky Way dust: for the alternate values ${\cal R} = 0$ or ${\cal R} = 4.1$, respectively, $\chi^2$ increases by $>$115 or $>$32.  


\begin{figure}
\epsscale{1.17}
\plotone{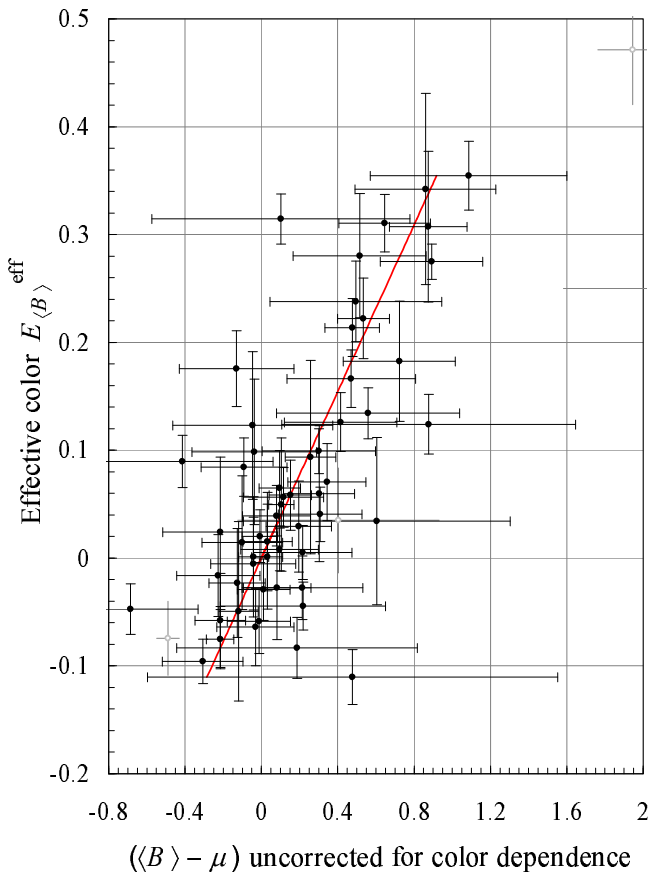}
\caption{Effective color $E^{\rm eff}_{\langle B \rangle}$ ($\equiv$ coefficient of ${\cal R}$, defined in equation (\ref{eqL})) {\it vs.}~residual from Hubble fit to $\langle B \rangle$ when the latter is left uncorrected by terms that are proportional to ${\cal R}$.  The line is a linear function of the ordinate determined by the global fit, from which SNe marked by open circles are omitted.} \label{Rcoeff}
\end{figure}


When the color is measured with less than perfect precision, the slope ${\cal R}$ will be underestimated.  In this analysis, the effects of finite color resolution were unfolded by two different methods described in Appendix \ref{appR}.  These methods yielded corrections to ${\cal R}$ of $+3.8\%$ and $+7.0\%$, respectively.  For comparison to other determinations of ${\cal R}$, which are likely to involve color resolutions similar to ours, we prefer to quote the uncorrected value ${\cal R} = 2.59 \pm 0.24$.  On the other hand, if our value of ${\cal R}$ is to be compared to other determinations for which the effects of color resolution already have been unfolded, we estimate our own correction for those effects to be given by the average of our two methods: $\Delta {\cal R}/{\cal R} = +0.054 \pm 0.027$, where a 50\% systematic error is assigned to the correction.

As an essential check, we have used the results of pseudoexperiments based on two independent Monte Carlo simulations to study the distributions both of pulls and residuals in ${\cal R}$ and other fit parameters.  In these pseudoexperiments, minimizing global $\chi^2$ to obtain ${\cal R}_{\rm fit}$ does reproduce on average the color-correction coefficient ${\cal R}_{\rm true}$ that was input to the simulation; measuring the variation of $\chi^2$ with ${\cal R}_{\rm fit}$ does predict the dispersion of ${\cal R}_{\rm fit} - {\cal R}_{\rm true}$ that is measured.  As is conventional, our global fit minimizes $\chi^2 = \delta^\dagger \, {\cal H} \, \delta$, where $\delta$ is a vector of differences and ${\cal H}$ is the information (inverse covariance) matrix.  This minimization problem belongs to the class in which both ${\delta}$ and ${\cal H}$ depend on the fit parameters; the latter dependence occurs because the uncertainties in Hubble residuals grow {\it e.g.}~as ${\cal R}$ increases.  For these Monte Carlo checks to be satisfied, we found it essential to allow ${\cal H}$ to vary continuously as the fit parameters were being optimized.  

We also tested an iterative stepwise fitting technique in which $\chi^2$ was minimized with ${\cal H}$ held fixed, then ${\cal H}$ was recalculated using the newly updated parameters, {\it etc}., until the iteration converged.  This caused our Monte Carlo checks to fail badly.  If applied to the real SN sample analyzed here, the stepwise technique would yield a value of ${\cal R}$ that is 10.2\% below that obtained with the validated fitting procedure.  Also, the measures of Hubble-line deviation to be discussed in \S \ref{wrms} would be marginally reduced (by $\approx$3\%).  This is understandable: slightly undercorrecting {\it e.g.}~for color variation can marginally improve the Hubble dispersion if the uncertainty associated with that correction can be reduced.  Nevertheless, all results presented in this paper are based on the Monte Carlo-validated fitting technique described in the previous paragraph, even at the expense of marginally larger Hubble dispersions.  We made this choice not only because the color- and decline-rate correction coefficients are physical properties of the SNe that we wish to report accurately, but also because in future campaigns, as the photometric errors {\it e.g.}~in color become relatively less important, the correction coefficients which minimize the Hubble dispersion will approach their true values.

\subsection{Upper Limit on Peculiar Velocity} \label{peculiar}

As described in \S \ref{noiseProp}, a single effective rms SN velocity $v_{\rm eff}$ is used to parametrize the effects both of peculiar velocity $v_{\rm pec}$ along the line of sight relative to the Hubble flow, and of any other non-photometric noise component that is common to $B_{\rm max}$ and $B_{BV}$.  In the Hubble fit for ${\langle B \rangle}$, $v_{\rm eff}$ is adjusted to maximize the probability density of $\chi^2$, yielding the best fit value $v_{\rm eff} = 382^{+60}_{-52}$ km s$^{-1}$.  As noted in \S \ref{noiseProp}, measuring $v_{\rm eff}$ yields an upper bound on SN peculiar velocity.  The corresponding 95\% confidence upper limit is $v_{\rm pec} < 486$ km s$^{-1}$.

In the literature, estimates are available for the peculiar velocities of individual SNe in the Local Group; applying corrections based on those estimates would improve significantly our Hubble fits.  However, our aim here is to use local SNe to characterize the peculiar velocities of distant SNe, for which such estimates generally are unavailable.  Therefore the results quoted above for $v_{\rm eff}$ and $v_{\rm pec}$ intentionally were left uncorrected for the estimated velocities, relative to the Hubble flow, of local SNe.


\begin{figure}
\epsscale{1.17}
\plotone{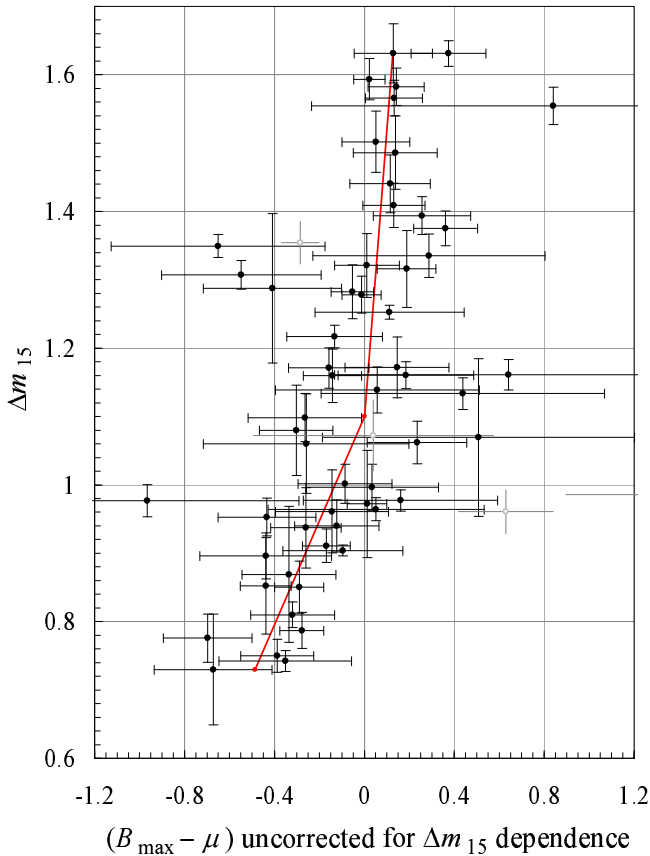}
\caption{Decline rate $\Delta m_{15}$ {\it vs.}~Hubble residual $(B_{\rm max} - \mu)$, which here is uncorrected for its dependence on $\Delta m_{15}$.  The line is the bilinear function of $\Delta m_{15}$ from the global fit, from which SNe marked by open circles are omitted.}\label{dogleg3}
\end{figure}


\section{RESULTS FOR $B_{\rm max}$ AND $B_{BV}$} \label{Bmax}

As emphasized in \S \ref{global}, the fit to $\Delta E = 0$ and the Hubble fit to ${\langle B \rangle}$ together comprise a global fit that fully determines the decline-rate and color corrections to be applied to $B_{\rm max}$ and $B_{BV}$ as well as to ${\langle B \rangle}$ and $\Delta E$.   In principle, no separate Hubble fits to $B_{\rm max}$ or $B_{BV}$ are required.  Figure \ref{dogleg3} plots the decline rate $\Delta m_{15}$ {\it vs.}~the Hubble residual for $B_{\rm max}$.  The corresponding decline-rate coefficients for $B_{\rm max}$ (and also for $B_{BV}$) are supplied in Table \ref{Dm153}.   There the symbols $\alpha$ and $\alpha'$ refer, respectively, to coefficients of $(\Delta m_{15} - 1.1)$ and $|\Delta m_{15} - 1.1|$; they are fully defined in Table \ref{Dm152}.  For alternate fits in which one or more decline-rate coefficients or combinations thereof are forced to vanish, the lower part of Table \ref{Dm153} exhibits the corresponding $\Delta \chi^2$ penalty.

As a check, independent Hubble fits to each of $B_{\rm max}$ and $B_{BV}$ were performed, allowing ${\cal R}$ and both bilinear decline-rate coefficients to seek optimum values independent of the global fit.  Relative to values of $\chi^2$ obtained for Hubble fits to $B_{\rm max}$ and $B_{BV}$ using parameters determined by the global fit, values of $\chi^2$ for these independent fits were smaller by less than 0.8 and 2.5, respectively.  This confirms that the global fit is a reasonable description of both $B_{\rm max}$ and $B_{BV}$.

\subsection{Decline-Rate Corrections to $B_{\rm max}$ and $B_{BV}$} \label{declineBmax}    

As a guide to the variety of numerical results from the global fit that are included in Table \ref{Dm153}, first we highlight those features which are firmly established by the data.  In particular, the results exhibited in rows 7 and 8 of Table \ref{Dm153}, derived from results of the fit to $\Delta E = 0$ summarized in Table \ref{Dm15}, emphasize that $B_{\rm max}$ and $B_{BV}$ cannot share the same decline-rate coefficients.  Because a kink in decline-rate correction at $\Delta m_{15} = 1.1$ is a feature of Type Ia SNe that may not be familiar to every reader, here we focus particularly on the coefficients $\alpha'$ of $|\Delta m_{15} - 1.1| \, $.  Because the tabulated values of $\alpha'_{\rm max}$ and $\alpha'_{BV}$ are nonzero, respectively, only by 2.7 and 1.2 standard deviations, a reader might imagine that the data marginally would allow both to vanish.  However, because the difference between those coefficients is nonzero at the $5.8 \sigma$ level, at least one of $B_{\rm max}$ and $B_{BV}$ must be described by a nonzero kink.  Other firmly established features include the need for a conventional decline-rate-correction term that is linear in $\Delta m_{15}$ for both $\langle B \rangle$ and $B_{\rm max}$ (rows 1 and 3 of the table).

Conversely, the data may allow certain coefficients in Table \ref{Dm153}, or combinations thereof, to vanish.  As one example, the result for $\alpha'_{\langle B \rangle}$ found in the second row of the table allows the decline-rate correction to $\langle B \rangle$ to be purely linear; that would correspond to the case in which the fit line in Fig.~\ref{dogleg2} is straight.  As a second example, fast-declining SNe (with $\Delta m_{15} > 1.1$) are corrected by a coefficient of $(\Delta m_{15} - 1.1)$ that is equal to $\alpha + \alpha'$.  As shown for corrections to $B_{\rm max}$ in the fourth-from-last row of Table \ref{Dm153}, $\alpha_{\rm max} + \alpha'_{\rm max}$ is only 1.5 standard deviations from zero.  Therefore the data do not conclusively reject the proposition that fast-declining SNe require no decline-rate correction whatsoever to $B_{\rm max}$; that would correspond to the case in which the top part of the ``dogleg'' line in Fig.~\ref{dogleg3} is exactly vertical.

Regarding the {\sc cmagic} luminosity $B_{BV}$, the third-from-last row of Table \ref{Dm153} examines the proposition that $B_{BV}$ requires no decline-rate correction for any SN: when two dof are removed by nulling both $\alpha_{BV}$ and $\alpha'_{BV}$, $\chi^2$ grows by 9.2 (confidence level 0.01).  Evidently the data disfavor this proposition.  More restrictively, slow-declining SNe (with $\Delta m_{15} < 1.1$) are corrected by a $(\Delta m_{15} - 1.1)$ coefficient equal to $\alpha - \alpha'$; the results in rows 5 and 6 of Table \ref{Dm153} clearly permit $\alpha_{BV} - \alpha'_{BV}$ to vanish.  Therefore the data do allow slow-declining SNe to remain uncorrected for decline rate if analyzed using the {\sc cmagic} method.  For color as well, some of these SNe might acceptably remain uncorrected: ``bluish'' SNe (with $\bv$ color ${\cal E} \la 0.1$) have a color correction to $B_{BV}$ that is no greater than $({\cal R}-\beta_{BV}) \, {\cal E} \approx (2.6-2) \, (0.1) \approx  0.06$ magnitudes.

\subsection{Allocating Intrinsic Noise to $B_{\rm max}$ or $B_{BV}$} \label{noiseAlloc}

As explained in \S \ref{noiseProp}, the intrinsic noise term $n_{\Delta E}$ that is required by the fit to $\Delta E = 0$ propagates into a noise term $n_{\langle B \rangle} = {\cal R} n_{\Delta E}/2$ in the Hubble fit for ${\langle B \rangle}$.  Table \ref{alloc} presents the $\chi^2$ for the Hubble fit to $B_{\rm max}$, the $\chi^2$ for the Hubble fit to $B_{BV}$, and the relative likelihood (proportional to the product of the two $\chi^2$ densities) for three scenarios:  (a) all of $n_{\langle B \rangle}$ is allocated to $B_{BV}$; (b) $1/\sqrt{2}$ of $n_{\langle B \rangle}$ is allocated to $B_{BV}$ and $1/\sqrt{2}$ to $B_{\rm max}$; (c) all of $n_{\langle B \rangle}$ is allocated to $B_{\rm max}$.  All of the fits that contribute to the table use the parameters of the global fit.  Scenario (a)\ is favored, so it is the basis for the global fit results that are presented in this paper.  Scenarios (b)\ and (c)\ are progressively less likely, by factors of 3.3 and 44, respectively; evidently scenario (b)\ is not strongly disfavored relative to (a).  Although the data do not conclusively reject scenario (c), it is unlikely that all of the noise in the difference between $B_{\rm max}$ and $B_{BV}$ is due to noise that is confined to $B_{\rm max}$.


\tabletypesize{\footnotesize}
\setlength{\tabcolsep}{5pt}
\begin{deluxetable}{lccc}
\tablewidth{0pt}
\tablecolumns{4}
\tableheadfrac{}
\tablecaption{Consequences of allocating the intrinsic noise in $\Delta B$ all to $B_{BV}$, all to $B_{\rm max}$, or equally to both.\label{alloc}}
\tabletypesize{}
\tablehead{\colhead{Quantity}
&\multicolumn{3}{c}{Noise allocation}
}
\startdata
\nodata&all to $B_{BV}$&equal&all to $B_{\rm max}$\\
\cutinhead{Conditions and Hubble fit results:}
Noise in $B_{\rm max}$ (mag)&0&0.052&0.074\tablenotemark{a}\\
Noise in $B_{BV}$ (mag)&0.074\tablenotemark{a}&0.052&0\\
$\chi^2$ ($B_{\rm max}$ residuals)&43.2&39.3&36.5\\
$\chi^2$ ($B_{BV}$ residuals)&57.8&66.1&80.2\\
Relative likelihood&1&1/3.3&1/44\\
\enddata
\tablenotetext{a}{For ($B_{\rm max}-B_{BV}$) the fit noise is $0.074 \pm 0.019$ mag.}
\end{deluxetable}


\subsection{Quantifying Deviations about the Hubble Line} \label{wrms}

From the distribution of a property $m$ of any statistically limited sample, estimating the variance of the measured variance in $m$ of that sample requires no assumption about the shape of the true $m$ distribution.  In part for that reason, the unweighted rms Hubble residual $\sigma$ (the square root of the variance) is a useful standard measure of SN uniformity.  A drawback of the rms is that, if included in the sample, SNe with unusually large uncertainties in $m$ can dominate this measure, pushing the rms upward.  This shortcoming may be addressed by use of the weighted rms (wrms), denoted by $\sigma_w$.  The square of $\sigma_w$ is the mean square deviation of $m$ from the function to which it is fit, weighted by the inverse square uncertainty assigned to $m$ for each SN.  (Equations defining both $\sigma$ and $\sigma_w$ are provided in Appendix \ref{appS}.)  These definitions take proper account of the reduction in dof due to the number of variable parameters in the fit function. 

While it is invariant to the overall scale of the uncertainties assigned to $m$, the wrms is fragile if the individual uncertainties are assigned inaccurately.  As a pathological example, if 100 identical SNe are observed with identical methods, and if the uncertainty in $m$ of the first SN erroneously is set to 1\% of the value assigned to the rest, the wrms underestimates the true rms typically by a factor of $\ga$7.
 
Both the rms and the wrms are highly sensitive to the presence or absence of a few SNe far out on the tails of the distribution.  This sensitivity is addressed here by introducing as a third measure of SN uniformity the median absolute residual ${\rm med}(|m_i-\mu(z_i)|)$, or ``core deviation''.  We scale this quantity by the factor 1.483 so that it is equivalent to the rms for a pure gaussian distribution in the limit $N/N_{\rm dof} \to 1$.     

\subsection{Hubble-line Deviations of $\langle B \rangle$, $B_{\rm max}$, and $B_{BV}$.} \label{Hubble}

The first three rows of Table \ref{HubDev} list the wrms, rms, and core deviations from the Hubble line for $\langle B \rangle$, $B_{\rm max}$, and $B_{BV}$, using the single set of correction coefficients obtained from the global fit.  To suppress the effects of peculiar velocity, only SNe with $z > 0.015$ are included in these averages (Fig.~\ref{hub} exhibits the sample's $z$ distribution).  We estimate the uncertainties in values of the weighted rms and core deviation to be similar to the relatively well-defined uncertainties in the values of the rms that appear in the table.  That these uncertainties are as large as their estimates is emphasized by the final row of Table \ref{HubDev}, which shows the result of substituting for the output of the global fit the output of a Hubble fit that was made only to $B_{BV}$.  Residuals from the $B_{BV}$-only fit are shown in Fig.~\ref{hub}.  In this fit, after adding 0.053 {mag} of noise in quadrature to the photometric errors in $B_{BV}$, we rejected as outliers five SNe chosen to minimize $\chi^2$ for the $B_{BV}$-only fit rather than for the global fit.  The resulting $\chi^2$ probability density was maximal.  Two SNe that had been excluded from the global fit, SN1992bp and SN1994ae (Fig.~\ref{circle}), were replaced as outliers by SN1995ac and SN2001V.  As is apparent, substituting the $B_{BV}$-only for the global fit perturbed the Hubble-line deviations for $B_{BV}$ by amounts comparable to other differences found in the table.  Weighing all of the information in Table \ref{HubDev}, it is evident that, with the available statistics, the Hubble-line deviations for all three measures of SN luminosity -- $\langle B \rangle$, $B_{\rm max}$, and $B_{BV}$ -- are comparable and of order $\approx$0.14 mag, and are not easily distinguished from each other.  Note that these deviations arise from the effects of photometric error and peculiar velocity as well as intrinsic noise.


\tabletypesize{\footnotesize}
\setlength{\tabcolsep}{2.4pt}
\begin{deluxetable}{ccccc}
\tablewidth{0pt}
\tablecolumns{5}
\tableheadfrac{}
\tablecaption{Weighted rms deviation, rms deviation, and core deviation (mag) from the Hubble line over the range $0.015 < z < 0.1$ for $\langle B \rangle$, $B_{\rm max}$, and $B_{BV}$ using the global fit and for $B_{BV}$ using a $B_{BV}$-only fit.\label{HubDev}}
\tabletypesize{}
\tablehead{\colhead{\begin{tabular}{c} Hubble \\ deviation \\ of \end{tabular}}
&\colhead{\begin{tabular}{c} Type \\ of fit \end{tabular}}
&\colhead{\begin{tabular}{c} Weight- \\ ed rms \end{tabular}}
&\colhead{rms}
&\colhead{\begin{tabular}{c} 1.483 $\times$ \\ median \\$|$residual$|$ \end{tabular}}
}
\startdata
$\langle B \rangle$&global&0.129&0.152$\pm$0.018&0.116  \\
$B_{\rm max}$&global&0.130&0.152$\pm$0.015&0.144  \\
$B_{BV}$&global&0.155&0.176$\pm$0.018&0.158  \\
$B_{BV}$&$B_{BV}$ only&0.132&0.144$\pm$0.015&0.120  \\
\enddata
\end{deluxetable}


\begin{figure}
\epsscale{1.17}
\plotone{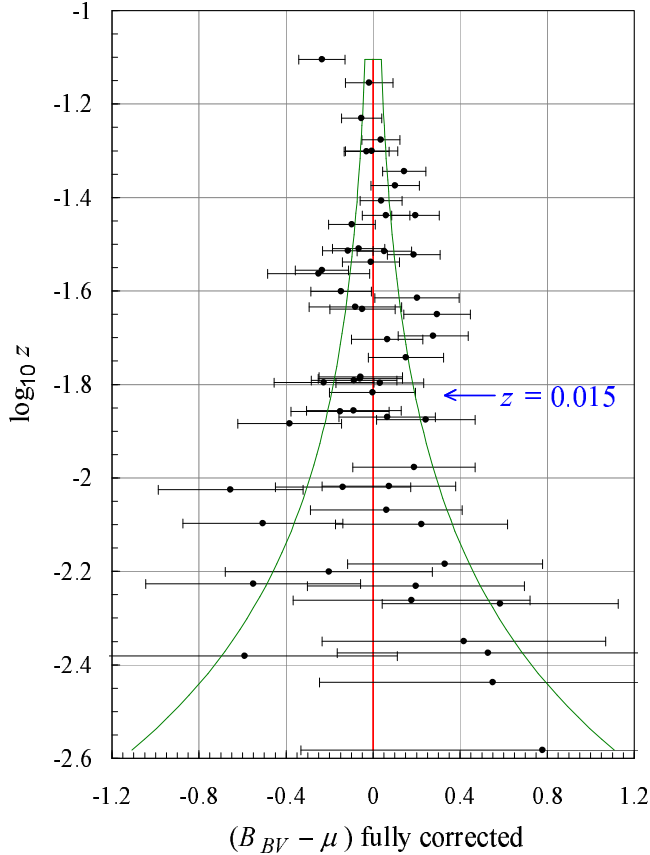}
\caption{Log$_{10}$(redshift $z$) {\it vs.}~{\sc cmagic} Hubble residual $(B_{BV} - \mu)$.  The points are residuals from the $B_{BV}$-only fit.  Five SNe excluded as outliers are suppressed.  The curve shows the $\pm 1 \sigma$ uncertainty in residual associated with the assumed rms line-of-sight peculiar velocity of 400 km/s relative to the Hubble flow.  Accounting for the dof removed by the corrections, the unweighted rms deviation from the fit (straight line) of the points with $z > 0.015$ is $0.144 \pm 0.015$ {mag} (Table \ref{HubDev}).}\label{hub}
\end{figure}


\section{CHARACTERISTICS OF THE BLUE RIDGE} \label{blue}

SNe that are considered to be part of the ``blue ridge'' are grouped in a region roughly defined as $|B_{\rm max} - V_{\rm max}| \la 0.1$; the word ``ridge'' refers to the contour of this group on a plot of color {\it vs.}~decline rate, as in Fig.~\ref{ridge}.  SNe in the blue ridge are thought largely to be free of extinction due to host-galactic dust \citep{Hatano:1998,Commins:2004}.  If this is so, sufficiently well-observed variation of SN color within the ridge area could reveal the distribution of intrinsic SN color.  Here we examine whether such observed color is correlated with decline rate $\Delta m_{15}$, and whether SN luminosities require correction for its variation.

\subsection{Decline-Rate Dependence of Ridgeline}\label{blueDm}

We have searched for evidence of a linear dependence upon $\Delta m_{15}$ of the blue ridgeline by making an unbinned maximum-likelihood fit of the data in Fig.~\ref{ridge} to a ``ridge'' color-distribution function (whose most probable color varies as ${\cal A} \, (\Delta m_{15}-1.1)$, where ${\cal A}$ is a fit constant), plus a $\Delta m_{15}$-independent ``tail'' color-distribution function that is nonzero only on the red side of the ridge peak.  (For the narrow purpose of this fit, the advantage of this simple color-distribution function over a more realistic model, such as that of \citet{Hatano:1998}, is that SNe far outside the ridgeline have smaller influence on the fit results.)  Within statistics ($\sigma({\cal A}) \approx 0.06$), we find no evidence for a nonzero value of the coefficient ${\cal A}$.  Because our blue ridge is consistent with being $\Delta m_{15}$-independent, the only variations with decline rate of our corrections for SN color are those represented by eqs.~(\ref{eqK}-\ref{eqL}).

\subsection{Do SNe in the Blue Ridge Require Color Correction?}\label{blueR}

If SNe that are clustered in the blue ridge are largely unextinguished by host-galactic dust, so that they exhibit mainly a spread in intrinsic color, any necessary color correction to their luminosities could depart substantially from corrections that normally are applied to extinction by dust.  In particular, the data might require little or no correction to luminosity for intrinsic color variation.  If this simple picture were correct, restricting analysis to SNe that are clustered in the blue ridge could substitute for implementing a color correction.

To test this simple model, we need to identify those SNe which truly belong to the blue ridge group.  To perform this identification while avoiding any chance of prejudice against the model, adventurously we allow the Hubble residuals of the SNe, as well as their colors, to play a role in the process of SN selection.  For simplicity and for ready comparison with other work, in this section we consider only the Hubble residuals for $B_{\rm max}$, and, in Hubble fits to $B_{\rm max}$, we allow all extraneous parameters including decline-rate coefficients to assume their best fit values independent of the results of any other fit.

Our scheme for identifying a group of SNe that have a relatively good chance of obeying the simple model described above is illustrated in the scatter plot shown in Fig.~\ref{cluster}.  There, for each SN, the abscissa is its pull in the Hubble fit to $B_{\rm max}$ assuming no color correction, {\it i.e.}~${\cal R} = 0$.  The ordinate is a ``color pull'' whose numerator is the excess of $B_{\rm max} - V_{\rm max}$ above a typical value of 0.02 {mag}, divided by the sum in quadrature of the photometric error and a characteristic intrinsic color rms of 0.03 {mag} taken from Phil99.  As in Fig.~\ref{circle}, SNe that lie within the circle have a sum of pulls squared that is less than 8.  These 41 SNe comprise the sample that we select to have a relatively good chance of obeying the ``${\cal R} = 0$'' simple model.  Among the SNe rejected by this cut are SN1993O, which is hosted by an elliptical galaxy; and SN1992bh and SN1995al, which belong to a range of observed color that we do find to be characteristic of the observed colors of SNe that are hosted by E, E/S0, and S0 galaxies.

It may appear na{\"i}ve to accept as many as 41 of the 56-SN fitted sample (Table \ref{cut}) into a sample to be tested for freedom from extinction.  However, the mean and rms $B_{\rm max}-V_{\rm max}$ for all 41 of these SNe (0.022 and 0.061, respectively) are comparable to those (0.022 and 0.062) of the sample of 13 SNe hosted by E, E/S0, or S0 galaxies, whose reddening would be expected to be less significant than for the balance of SNe that occurred in spiral galaxies.


\begin{figure}
\epsscale{1.17}
\plotone{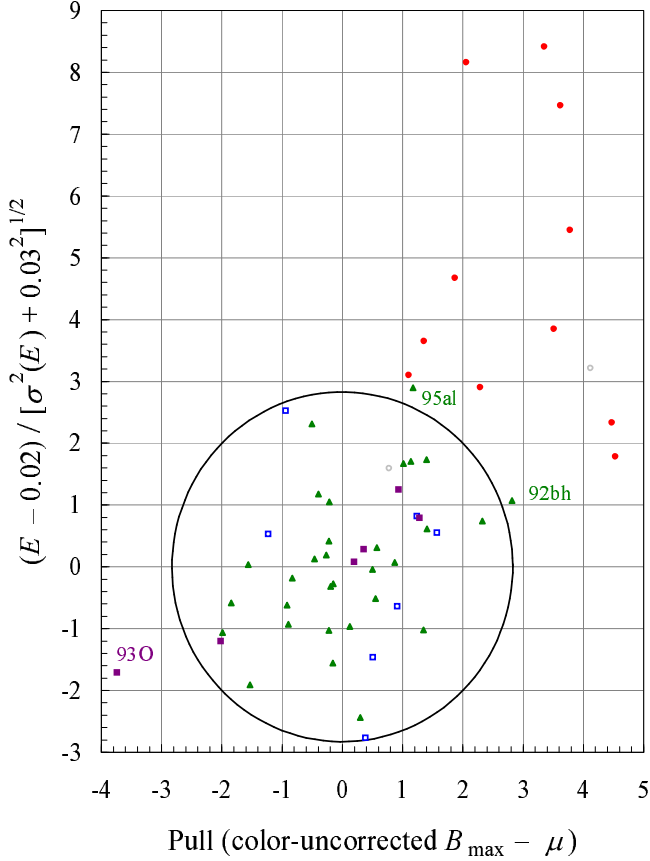}
\caption{``Color pull'' (defined in the text) {\it vs.}~pull in a color-uncorrected Hubble fit to $B_{\rm max}$, for SNe in five classes depicted by the same symbols (and colors in electronic edition) that are used in Fig.~\ref{ridge}.  SNe that are clustered within the circle of radius $\sqrt{8}$ are identified as having a relatively good chance of not requiring a correction for color variation. } \label{cluster}
\end{figure}



\tabletypesize{\footnotesize}
\setlength{\tabcolsep}{4pt}
\begin{deluxetable}{lccc}
\tablewidth{0pt}
\tablecolumns{4}
\tableheadfrac{}
\tablecaption{Details of SN selection and results of fits that reject a simple model in which SNe that are clustered within the blue ridge require no color correction.\label{signif}}
\tabletypesize{}
\tablehead{\colhead{\begin{tabular}{c} Quantity \end{tabular}}
&\multicolumn{2}{c}{Inside cluster}
&\colhead{\begin{tabular}{c} Outside \\ cluster \end{tabular}}
}
\startdata
Pull$^2$ (color) + pull$^2$ (Hubble)&\multicolumn{2}{c}{$<8$}&$>8$\\
SNe with E, E/S0, or S0 hosts&\multicolumn{2}{c}{$12$}&\phn1\\
SNe with late-type hosts&\multicolumn{2}{c}{$29$}&14\\
\cutinhead{Conditions and results for Hubble fits to $B_{\rm max}$:}
${\cal R}$ (fixed in fit to 41 SNe)&2.59\tablenotemark{a}&0&\nodata\\
$\chi^2$ (of fixed-${\cal R}$ fit to 41 SNe)&31.8&45.6&\nodata\\
Relative likelihood&1&0.001&\nodata\\
\enddata
\tablenotetext{a}{From global fit to 56 SNe, ${\cal R} = 2.59 \pm 0.24$.}
\end{deluxetable}


The final step in testing the simple ``${\cal R} = 0$'' model is to subject this 41-SN subsample to two alternative Hubble fits:  one with ${\cal R} = 0$ and the other with ${\cal R} = 2.59$, the result of the global fit described in \S \ref{colorHub}.  The details of the SN selection and fit results are shown in Table \ref{signif}.  In the top half of the table, the color and Hubble pulls are those described above.  In the bottom half, the values of $\chi^2$ are for Hubble fits to $B_{\rm max}$ with ${\cal R}$ fixed to either of the two indicated values.  Even for the restricted subsample of SNe that is fit, $\chi^2$ for the ${\cal R} = 2.59$ fit is substantially smaller than for ${\cal R} = 0$, by a difference of 13.8.  Therefore, with 99.9\% confidence, the data reject the hypothesis that SNe in the blue ridge group require no color correction.  Whether this correction is for intrinsic color or for dust, it is needed.  Had we selected the test sample less aggressively -- for example, had we declined to include among the selection criteria the sizes of the ${\cal R} = 0$ Hubble residuals themselves -- this conclusion would only be strengthened.

\subsection{Color Corrections to SNe with E, E/S0, or S0 Hosts}\label{elip}

Of the 41-SN subsample chosen for the test just described, 12 are hosted by elliptical, E/S0, or S0 galaxies.  A thirteenth, SN1993O, has an elliptical host but was rejected from that subsample because, if uncorrected for color, it exhibited a large ${\cal R} = 0$ Hubble $|$residual$|$ (Fig.~\ref{cluster}).  One might well ask whether these SNe, hosted by relatively ancient galaxies with little remaining dust, also require a nonzero color coefficient ${\cal R}$.

From the data considered in this paper, unfortunately, the answer is ambiguous.  If SN1993O continues to be omitted from the Hubble fits, the remaining 12 SNe that are hosted by E+E/S0+S0 galaxies show little preference for either of the two values of ${\cal R}$ over the other.  On the other hand, if SN1993O is included, ${\cal R} = 2.59$ is preferred over ${\cal R} = 0$.  Because a robust result cannot hinge on including or excluding a single SN, a definite conclusion on the color correction of SNe in elliptical, E/S0, and S0 galaxies is not reached here.  

\section{SUMMARY AND DISCUSSION} \label{summary}

Applying both a substantially extended light-curve-fitting algorithm and the {\sc cmagic} prescription of Wang03 to published $B$ and $V$ fluxes from a systematically chosen sample of nearby SNe, we have compared the outputs $B_{\rm max}$ and $B_{BV}$ of these two types of fit both to each other and to Hubble's-law expectation.  After applying standard corrections for observed SN color, but before applying decline-rate corrections, the difference $B_{\rm max} - B_{BV}$ reduces to ${\cal R} \, \Delta E$, where ${\cal R}$ is a generalized linear color-correction coefficient of the type measured by \citet{Tripp:1998}; and $\Delta E \equiv {\cal E} - (B_{\rm max} \!-\! V_{\rm max})$ is the difference between standard measures of $\bv$ color used, respectively, by the {\sc cmagic} and light-curve fit methods.  

After $\Delta E$ is corrected for SN decline rate $\Delta m_{15}$, close agreement in the two measures of $\bv$ color is achieved: their mutual rms is only $0.053 \pm 0.006$ mag.  The required correction is compatible with a bilinear but not a linear $\Delta m_{15}$ dependence (Fig.~\ref{dogleg}); the nonlinear term is highly significant (5.8$\sigma$).  Also, to bring the $\chi^2$ for the fit to $\Delta E = 0$ down to an acceptable level, the propagated photometric errors in $\Delta E$ must be augmented in quadrature by an intrinsic noise term $n_{\Delta E} = 0.029 \pm 0.007$ mag rms.

In {\sc cmagic} plots of $B$ {\it vs.}~$\bv$ shown in their Figs.~(1-4), Wang03 distinguished between SNe that did or did not show a ``bump'' near maximum light.  As is evident from those plots, SNe which did exhibit a bump had a $B$ flux at maximum that was larger than would have been obtained by extrapolating to the same epoch the linear fit to $B$ {\it vs.}~$\bv$ that was made in the post-maximum {\sc cmagic} region.  Therefore the prominence of a bump in the {\sc cmagic} plot is intimately related to the difference between $B_{\rm max}$ and $B_{BV}$, and therefore to the size of $\Delta E$.  Wang03 noted further that SNe which exhibited a bump tended to be slow decliners, with low $\Delta m_{15}$.  The continuous dependence of $\Delta E$ upon $\Delta m_{15}$ depicted in our Fig.~\ref{dogleg} suggests that SNe may not be bifurcated into bumpless and bumpful subsets; rather, their systematic differences $B_{\rm max} - B_{BV}$ may be described by a continuous (but nonlinear) function of SN decline rate that reaches its nadir for the slowest decliners, as would be expected from Wang03's observations.

Turning to results from our light-curve fits, we have studied the group of SNe which inhabit the ``blue ridge'' region $|B_{\rm max} - V_{\rm max}| \la 0.1$.  In particular, on a plot of $B_{\rm max} \!-\! V_{\rm max}$ {\it vs.}~$\Delta m_{15}$ (Fig.~\ref{ridge}), we sought but did not find evidence for a linear $\Delta m_{15}$-dependence of the $\bv$ color about which this group is clustered:  within an uncertainty of $\approx$0.06 the best fit $\Delta m_{15}$ coefficient ${\cal A}$ is consistent with zero.  Taking ${\cal A} = 0$ as a fair representation of the blue ridge topology on this plot, and using the result of our best fit to the $\Delta m_{15}$-dependence of $\Delta E$, we can deduce the topology of the blue ridge on a plot of {\sc cmagic} color ${\cal E}$ {\it vs.}~$\Delta m_{15}$.  On such a plot, for $\Delta m_{15} < 1.1$ the color of the ridgeline would redden in proportion to $\Delta m_{15}$ with a slope $\alpha_{\Delta E} - \alpha'_{\Delta E} = 0.53$ (Table \ref{Dm15}); at higher decline rates it would deredden with a much gentler slope $\alpha_{\Delta E} + \alpha'_{\Delta E} = -0.07$.  On their plot of ${\cal E}$ {\it vs.}~$\Delta m_{15}$, Wang03 fit the blue ridge to a linear dependence on $\Delta m_{15}$ over its full range, obtaining a slope equal to $0.25 \pm 0.04$, in agreement with our average slope $\alpha_{\Delta E} = 0.23 \pm 0.03$.  To sum up, when measured at peak luminosity the blue ridge color shows little dependence on decline rate; but when the color assigned to SNe in the ridge is strongly influenced by post-maximum values of $\bv$, as occurs for the {\sc cmagic} color ${\cal E}$, the ridge color tends to redden as the decline rate steepens, particularly for slow decliners.

From their study of $B_{\rm max} \!-\! V_{\rm max}$ {\it vs.}~$\Delta m_{15}$, Phil99 reported a blue ridge slope ${\cal A} = 0.114 \pm 0.037$ for SNe that were preselected to exhibit low $\bv$ color excess in the nebular phase (30-90 days after maximum light).  This $\Delta m_{15}$ dependence is intermediate between the dependences we observe, with no comparable preselection, for $B_{\rm max} \!-\! V_{\rm max}$ and for ${\cal E}$.  Our $\Delta m_{15}$ dependence for $B_{\rm max} - V_{\rm max}$ is not significantly different from that of Phil99.
  
A Hubble fit to the average of $B_{\rm max}$ and $B_{BV}$, when combined with the fit to $\Delta E = 0$, is equivalent to a global fit that specifies the color and decline-rate corrections for both $B_{\rm max}$ and $B_{BV}$.  It yields the $\bv$ color-correction coefficient ${\cal R} = 2.59 \pm 0.24$.  This value, which parametrizes the effects both of extinction by host-galactic dust and of intrinsic SN color variation, lies well below the coefficient $R_B \approx 4.1$ that conventionally is used for Milky Way dust.  The fit result would increase by $\Delta {\cal R}/{\cal R} = +0.054 \pm 0.027$ if we were to unfold the effects of finite color resolution, and it would diminish by $\Delta {\cal R}/{\cal R} = -0.102$ if we were to bias it by performing a stepwise rather than a continuous $\chi^2$ minimization, as is discussed in \S \ref{colorHub}.  Our measurement of ${\cal R}$ is consistent with the values $2.09 \pm 0.38$; 2.44; $2.46 \pm 0.46$; and $2.19 \pm 0.33$ obtained, respectively, by \citet{Tripp:1998}; \citet{Tripp:1999}; \citet{Parodi:2000}; and \citet{Guy:2005}.  (Note, however, that \citet{Tripp:1999} and \citet{Guy:2005} report having performed a stepwise $\chi^2$ minimization.)  Our result for ${\cal R}$ is incompatible with the values 1.71 and 3.5 used, respectively, by \citet{Saha:1999} and \citet{Altavilla:2004}.

That ${\cal R}$ is significantly below the canonical value of $\approx 4.1$ for Milky Way dust is a real puzzle that we would like to understand.  Very recently, \citet{Wang:2005} advanced an explanation of the low value of $R_B$ that involves scattering by dust clouds located in the circumstellar (CS) environment.  From CS dust at the radius ($10^{16}$ cm) modeled in that paper, scattered light would be delayed significantly relative to unscattered light emitted directly by the SN, distorting the evolution of its observed luminosity and color.  In the present paper, where statistical analysis of our 56-SN sample yields an example of the low value of $R_B$ which CS dust purports to explain, similar statistical techniques may be used to search for such distortions.  For example, larger amounts of CS dust could make the decline rate parameter $\Delta m_{15}$ appear smaller as a result of the delay in scattered light, causing a negative correlation between color and decline rate.  

Pursuing this point, Fig.~3(a-b) of \citep{Wang:2005} illustrates the effects of that paper's CS dust model on a single SN.  Its $\bv$ color is $E \approx -0.03$ when unextinguished, {\it vs.}~$E \approx 0.24$ after absorption and scattering by CS dust.  Taking into account the change in time of $B_{\rm max}$, $\Delta m_{15}$ drops from $\approx 0.95$ without extinction to $\approx 0.70$ after CS dust extinction, yielding the quotient $Q \equiv \Delta (\Delta m_{15}) / \Delta E \approx -0.25/0.27 \approx -0.92$.  If one interprets $Q$ as the coefficient of a linear dependence upon $E$ of (observed) $\Delta m_{15}$, for these variables one predicts a Pearson correlation coefficient ${\rm R}_{\rm pred} = Q \, \sigma(E)/\sigma(\Delta m_{15}) \approx -0.52 \pm 0.08$, where the error arises from uncertainties on the standard deviations $\sigma$ of $E$ and $\Delta m_{15}$.  For comparison we examined the 43 late-type SNe shown by closed triangles and closed circles in Fig.~\ref{ridge}, whose correlation between $E$ and $\Delta m_{15}$ is ${\rm R}_{\rm obs} = 0.07 \pm 0.16$.  Barring a $\ga 3 \sigma$ fluctuation,  ${\rm R}_{\rm obs}$ would have been sensitive statistically to the CS dust prediction.  However, if part of the spread in $E$ were due to variation in {\sl intrinsic} color $E_{\rm intr}$, and if true $\Delta m_{15}$ were {\sl positively} correlated with $E_{\rm intr}$ (as suggested by Fig.~5(a) of Phil99), a negative correlation due to CS dust could have been cancelled, accounting for our low value of ${\rm R}_{\rm obs}$.  We also searched for a more general dependence of $\Delta m_{15}$ upon $E$ by dividing the same 43-SN sample into reddened {\it vs.}~unreddened subsamples (filled circles {\it vs.}~filled triangles in Fig.~\ref{ridge}).  Their difference in $\langle E \rangle$ matched $\Delta E$ of the single SN modeled by \citep{Wang:2005}.  Again, at a comparable level of sensitivity, we did not observe the relative drop in $\langle \Delta m_{15} \rangle$ predicted for the reddened subsample.  As models of circumstellar dust around SNe are refined, statistical constraints of the type described here should prove valuable.  

Resuming the summary of this paper's results, we return briefly to SNe that inhabit the blue ridge.  There we selected a subsample of SNe that offer a relatively good chance of not requiring any color correction.  Notwithstanding the possible bias in their selection, in separate fits to the Hubble residuals of $B_{\rm max}$ these SNe prefer ${\cal R} = 2.59$ to ${\cal R} = 0$ with a likelihood ratio of 1000:1.

The nonlinear decline-rate correction required to reconcile the $B_{\rm max} \!-\! V_{\rm max}$ and {\sc cmagic} colors forces $B_{\rm max} - B_{BV}$, which is proportional to the color difference, to exhibit the same nonlinearity.  Therefore, with the same 5.8$\sigma$ significance, a nonlinear decline-rate dependence must characterize either $B_{\rm max}$, $B_{BV}$, or both.  When the global fit is used to apportion that nonlinearity between $B_{\rm max}$ and $B_{BV}$, the nonlinear term is insignificant in $B_{BV}$ (1.2$\sigma$) and is only marginally significant in $B_{\rm max}$ (2.7$\sigma$).  Therefore we conclude only that the required nonlinearity is unlikely to be confined fully to $B_{BV}$.  Accepting our best-fit nonlinear parametrization of the $\Delta m_{15}$-dependence of $B_{\rm max}$ (Fig.~\ref{dogleg3}), over the indicated range of $\Delta m_{15}$ its maximum $|$deviation$|$ from the best-fit {\it linear} parametrization is 0.13 mag.  Therefore the need for a nonlinear decline-rate correction should not seriously challenge the validity of cosmological results based on past high-$z$ SN observations, upon which scant photon statistics have imposed other uncertainties that typically are greater.  

Phil99 introduced a decline-rate correction that was quadratic in $\Delta m_{15}$.  In Hubble fits to the more limited data then available, their quadratic term was not significant (0.9$\sigma$), and the sign of their best fit nonlinearity was such that faster decliners require steeper corrections.  In the present analysis, slower decliners require steeper corrections (Fig.~\ref{dogleg3}).  However, it is unreasonable to compare Fig.~\ref{dogleg3} directly with the results of Phil99; their $B_{\rm max}$ magnitudes were corrected for host-galactic reddening using a color coefficient $R_B \approx 4.1$ that is characteristic of Milky Way dust, whereas ours were corrected for host-galactic reddening plus intrinsic SN color variation using the coefficient ${\cal R} = 2.59 \pm 0.24$ that we measure in the current analysis.  We note that, for slow decliners, our best fit $\Delta m_{15}$ correction to the {\sc cmagic} output $B_{BV}$ is consistent with nil.

The proportionality of $B_{\rm max} - B_{BV}$ to the color difference $\Delta E$ causes the measured noise $n_{\Delta E}$ to map into a mutual intrinsic scatter $n_{\Delta B} = 0.074 \pm 0.019$ {mag} in $B_{\rm max}$ {\it vs.}~$B_{BV}$.  The values of $\chi^2$ for Hubble fits to $B_{\rm max}$ and $B_{BV}$ are such that the maximum likelihood occurs when all of $n_{\Delta B}$ is confined to $B_{BV}$.  However, the data do not significantly prefer that solution to one in which, for example, $1/\sqrt{2}$ of $n_{\Delta B}$ is apportioned to each luminosity measure.  At the level of a 44:1 likelihood ratio, it is unlikely that all of $n_{\Delta B}$ is confined to $B_{\rm max}$.  Even the full 0.074 mag is not a high level of scatter compared to the additional errors of $\approx 0.15$-$0.18$ mag often added in quadrature to photometric errors on the luminosities of high-$z$ SNe.  

Of course, additional intrinsic scatter that similarly affects both $B_{\rm max}$ and $B_{BV}$ may be present.  Lacking the statistical power to separate it from the effects of peculiar velocity, we lump those noise sources into an effective rms SN velocity $v_{\rm eff} = 382^{+60}_{-52}$ km s$^{-1}$.  This places a 95\% confidence upper limit of 486 km s$^{-1}$ on the true rms line-of-sight peculiar velocity of nearby SNe relative to the Hubble flow.  Most recent SN analyses employ values of the rms peculiar velocity that are below this limit; a particular exception is the recent SN calibration paper of \citet{WangX:2005}, whose analysis used a value of 600 km s$^{-1}$.  An rms peculiar velocity of 500 km s$^{-1}$, barely above our limit, was assumed by \citet{Tonry:2003} and \citet{Barris:2004}.

Applying three different measures of dispersion, we find that $B_{\rm max}$, $B_{BV}$, and their unweighted average are scattered about the Hubble line by amounts that are comparable and of order $\approx 0.14 \pm 0.02$ mag (for $0.015 < z < 0.1$), and that are not easily distinguished from each other (Table \ref{HubDev}).  The uncertainties inherent in such measures are substantial, not only from random fluctuations but also from systematic variations in identification and treatment of outliers and in other methodologies.  We urge caution in interpreting these and other estimates of the uniformity of Type Ia SNe.

Our priorities for extension of the work described in this paper include achieving a better understanding of correlations among the published uncertainties in raw flux measurements made on individual SNe, and putting that understanding to use in further-refined light-curve fits.  In particular, we look forward to better quantification of the systematic errors that are assigned to the fit outputs.

\acknowledgments

We thank Weidong Li for providing photometric data for SN2002el prior to publication.  This work was supported by the Director, Office of Science, of the U.S.~Department of Energy under Contract No.~DE-AC02-05CH11231.

\begin{appendix}

\section{STATISTICS} \label{appS}

In the Introduction, reference is made to the possibility of a general solution to the problem of estimating the deviation $\Delta$ of a single SN from a fiducial absolute SN magnitude, using a set ${\bf x} \equiv \{x_j\}$ (1$\le$$j$$\le$$n$) of its measured parameters.  Given a training set consisting of an ensemble of $N$ SNe in the nearby Hubble flow whose corresponding deviations $\Delta_i$ (1$\le$$i$$\le$$N$) are deduced from their redshifts, an example of such a general solution is provided by the ${\vec \alpha}${\sc pde} recipe of \citet{Knuteson:2002}.  A covariance matrix
\begin{equation}
\Sigma_{kl} = {1 \over N}\sum_{i=1}^N
  \left( ({\bf v}_i)_k - {\bar{\bf v}}_k \right) 
  \left( ({\bf v}_i)_l - {\bar{\bf v}}_l \right) \label{eqa}
\end{equation}
is constructed, where ${\bf v}_i \equiv ({\bf x}_i,\Delta_i)$, ${\bar{\bf v}}_i$ is the mean ${\bf v}_i$, and $k,l$ index the ($n$+1) components of ${\bf v}$.  The best estimate of $\Delta$ maximizes the joint probability density
\begin{equation}
p({\bf x},\Delta) = 
  {{1} \over {N(\sqrt{2\pi}h)^{(n+1)} \det^{1/2}(\Sigma)}}
  \sum_{i=1}^N 
  \exp{\left(-{({{\bf v}-{\bf v}_i)^\dagger \Sigma^{-1} ({\bf v}-{\bf v}_i)} 
              \over {2 h^2}}\right)} \; , \label{eqb}
\end{equation}
where $h$ is a smoothing parameter determined by $N$ and $n$.  

In \S \ref{wrms}, reference is made to the unweighted rms Hubble residual $\sigma$.  It is defined by
\begin{equation}
\sigma^2 \equiv {{1} \over {N_{\rm dof}}} 
 \sum_{i=1}^N {(m_i-\mu(z_i))^2}  \; , \label{eqQ}
\end{equation}
where $N_{\rm dof}$ is the number $N$ of SNe with corrected magnitude $m_i$ at redshift $z_i$, less the number of parameters in the fit function $\mu(z)$.  Reference is made also to the weighted rms (wrms) Hubble residual $\sigma_w$.  It is defined by
\begin{equation}
\sigma^2_w \equiv {{N} \over {N_{\rm dof}}} 
 \sum_{i=1}^N {{(m_i-\mu(z_i))^2} \over {\sigma_i^2}} \Big/ 
 \sum_{i=1}^N {{1} \over {\sigma_i^2}} \; , \label{eqR}
\end{equation}
where $\sigma_i$ is the total uncertainty in $m_i$, taking into account photometry, intrinsic noise, and peculiar velocity.

\section{LIGHT CURVE FIT DETAILS} \label{appLC}

In \S \ref{LCfits}, reference is made to the ``stretched'' SN phase $t^*$ introduced in equation (\ref{eqc}).  We used the function
\begin{equation}
t^*(t; s_0,\tau_0) = t \left( 1 + ({{1} \over {s_0}} - 1) \, 
  {{1 + (t/\tau_0)^2} \over {1 + (t/\tau_0)^6}} \right) \; . \label{eqStretch}
\end{equation}

Also in \S \ref{LCfits}, reference is made to the fact that our light-curve fits to a minority of SNe employed six parameters (including the delayed parameters $\epsilon_0$ and $t_d$ in equation (\ref{eqc})), rather than four.  As is seen in Table \ref{table1a}, 18 of 61 fitted SNe fell into that category.  Of those, 16 are the subset of the 61 SNe that were identified spectroscopically to be SN1991T-like.  As well, adding two parameters improved the quality of fit to SN1996X and SN1998bu.  Figure \ref{fit95bd} exhibits a sample 6-parameter light-curve fit.

In \S \ref{fitPhoto} it is asserted that the quantity $\Sigma$ (equation (\ref{eqEF})) roughly tracks the photometric error on $\Delta E \equiv {\cal E} - E$.  From the definitions of ${\cal E}$ and $E$ in \S \ref{fitPhoto}, 
\begin{equation}
\beta_{BV} \Delta E = \beta_{BV} V^{\rm raw}_{\rm max} -     
  (\beta_{BV}-1) B^{\rm raw}_{\rm max} - B^{\rm raw}_{BV}
  \; . \label{eqSigma}
\end{equation}
After approximating $\beta_{BV} \approx 2$, neglecting uncertainties in corrections for Milky Way extinction, and taking the three raw quantities in equation (\ref{eqSigma}) to be uncorrelated (see \S \ref{nondeg}), the photometric error on its right-hand side reduces to $\Sigma \, $.  

Also in \S \ref{fitPhoto}, reference is made to adjustment of the photometric fit errors that are flagged by the footnote in Table \ref{table1a}.  These adjustments were occasioned by the detailed graphical inspections of fit results noted in \S \ref{fitPhoto}.  When the result of a photometric fit showed a systematic deviation from measured fluxes, usually owing to rigidity of the light-curve template, we left all best-fit values untouched but perturbed upward the error assigned to the fit parameter, {\it e.g.}~$B^{\rm raw}_{\rm max}$, to bring it into $\la 2 \sigma$ agreement with the data.  No account was taken of the implications for Hubble fits.  When the error in $B^{\rm raw}_{\rm max}$ was scaled, so was the error in $\Delta m_{15}$.  Overall, 13\% of the errors quoted in Table \ref{table1a} were adjusted.     


\begin{figure}
\epsscale{0.49}
\plotone{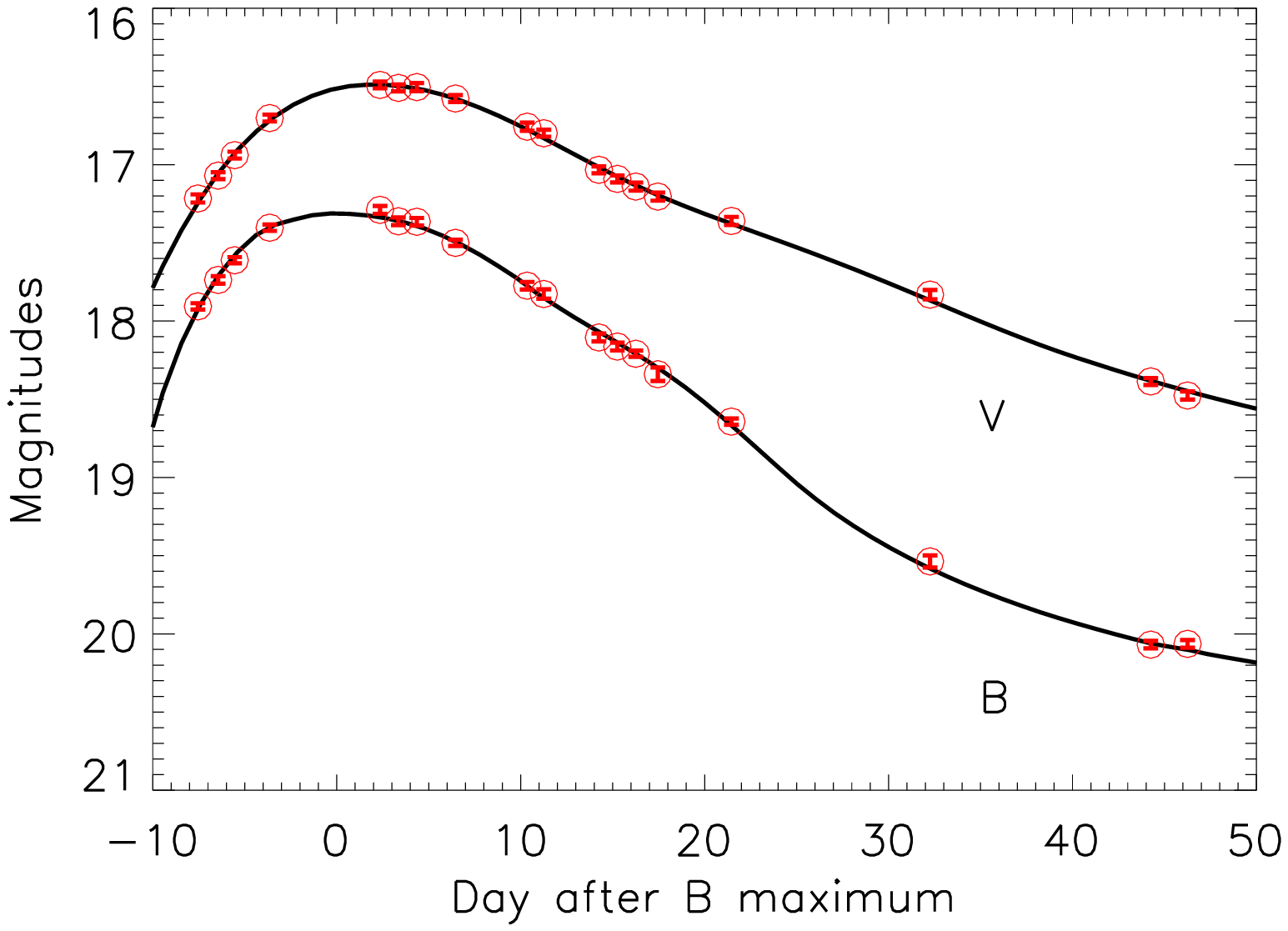}
\caption{Six-parameter light-curve fits to $B$ and $V$ {\it vs.}~rest-frame epoch for the highly extinguished Type Ia SN 1995bd.  Measured data points are shown by circles, with error bars indicating the 1$\sigma$ uncertainties.} \label{fit95bd}
\end{figure}


\section{EXAMPLE OF $B_{\rm max} \equiv B_{BV}$} \label{appD}

Under the special conditions enumerated in \S \ref{nondeg}, $B_{\rm max}$ and $B_{BV}$ are fully correlated.  As a simple example that satisfies these conditions, suppose that only three measurements $\{B_1,B_2,B_3\}$ of raw $B$ magnitude are made.  In the {\sc lc} fit to the $B$ band, after one fixes $\tau_0 = \infty$ and $\epsilon_0 = 0$ in equation (\ref{eqc}) and optimizes the three remaining parameters $\Gamma_0$, $t_0$, and $s_0$, normally one obtains a best fit light curve that passes exactly through the three measured $B$ points.  Suppose further that only a single measurement $V_3$ of $V$ magnitude is made, at the same epoch $t_3$ as $B_3$.  In the {\sc lc} fit to the $V$ band, $t_0$ and $s_0$ must then be determined by their $B$-band values, while, in this limiting case, only $\Gamma_0$ may be optimized.  Again the best fit light curve passes through the single measured $V$ point.  Within an additive constant, using the fact that all residuals vanish, 
\begin{mathletters}
\begin{eqnarray}
B^{\rm raw}_{\rm max} &=& B_3 - m_B(T_3) \; ; \\ \label{eqd}
V^{\rm raw}_{\rm max} &=& V_3 - m_V(T_3) \; ; \\ \label{eqe}
E &=& B_3-V_3  - (m_B(T_3)-m_V(T_3)) \; , \label{eqf}
\end{eqnarray}
\end{mathletters} 
where $m_{B,V}$ are the template magnitudes and $T_3 \equiv (t_3-t_0)/s_0$.  If $t_3$ lies within the {\sc cmagic} linear region, one may fix $\beta_{BV}$ at a representative value $\beta_0$ and use the single pair $\{B_3,V_3\}$ to determine $B^{\rm raw}_{BV}$;  within a constant,
\begin{mathletters}
\begin{eqnarray}
B^{\rm raw}_{BV} &=& \beta_0 V_3 - (\beta_0-1) B_3 \; ; \\ \label{eqg}
{\cal E} &=& B_3-V_3 - \beta_0^{-1} \, m_B(T_3)\; . \label{eqh}
\end{eqnarray}
\end{mathletters}
Forming $\Delta E \equiv {\cal E} - E$ and taking its differential,
\begin{equation}
{\rm d}(\Delta E) = {\rm d}(m_B(T_3)-m_V(T_3)) 
  - \beta_0^{-1} \, {\rm d}m_B(T_3) \; . \label{eqj}
\end{equation}
If the templates are ``magic'' -- that is, if they satisfy the condition ${\rm d}m_B/{\rm d}(m_B \!-\! m_V) = \beta_0$ in the {\sc cmagic} linear region -- the differential ${\rm d}(\Delta E)$ vanishes:  apart from an additive constant, the two color measures $E$ and ${\cal E}$ are equivalent, and so are $B_{\rm max}$ and $B_{BV}$ (eqs.~(\ref{eqC}) and (\ref{eqE})).  Therefore, if the data are minimal and the templates are magic, the maximum-luminosity and {\sc cmagic} analyses are mutually equivalent.

\section{$K$ CORRECTION DETAILS} \label{appK}

A brief description of our approach to $K$ corrections is found in \S \ref{Kcorrec}.  Table \ref{Kcorr} exhibits the difference between $K$-corrected and $K$-uncorrected $B_{\rm max}$, $V_{\rm max}$, $\Delta m_{15}$, $B_{BV}$, and $\beta_{BV}$, for four selected SNe whose redshifts and colors $E \equiv B^{\rm raw}_{\rm max} - V^{\rm raw}_{\rm max}$ are provided in the final two rows.  SN1992bp was selected because it has the largest redshift in the sample; SN1992al differs from it primarily in redshift.  SN1995ak and SN2000cn were included because they differ from each other primarily in color.


\tabletypesize{\footnotesize}
\setlength{\tabcolsep}{4pt}
\begin{deluxetable}{crrrr}
\tablewidth{0pt}
\tablecolumns{5}
\tableheadfrac{}
\tablecaption{$K$ corrections to $B_{\rm max}$, $V_{\rm max}$, $\Delta m_{15}$, $B_{BV}$ (mag), and $\beta_{BV}$, for four SNe with redshifts and colors listed in the last two rows.\label{Kcorr}}
\tabletypesize{}
\tablehead{\colhead{\begin{tabular}{c} $K$ correction \\ for SN \\ to quantity \end{tabular}}
&\colhead{1992bp}
&\colhead{1992al}
&\colhead{1995ak}
&\colhead{2000cn}
}
\startdata
$B_{\rm max}$&$-0.027$&$0.009$&$0.008$&$-0.013$\\
$V_{\rm max}$&$0.043$&$0.012$&$0.009$&$-0.005$\\
$\Delta m_{15}$&$-0.118$&$-0.019$&$-0.049$&$-0.093$\\
$B_{BV}$&$0.223$&$0.022$&$0.037$&$0.040$\\
$\beta_{BV}$&$-0.226$&$-0.019$&$-0.038$&$-0.038$\\
\cutinhead{Characteristics of $K$-corrected SNe:}
$z$&$0.0786$&$0.0135$&$0.0230$&$0.0232$\\
$E$&$-0.050$&$-0.051$&$0.000$&$0.202$\\
\enddata
\end{deluxetable}


\begin{figure}
\epsscale{0.5}
\plotone{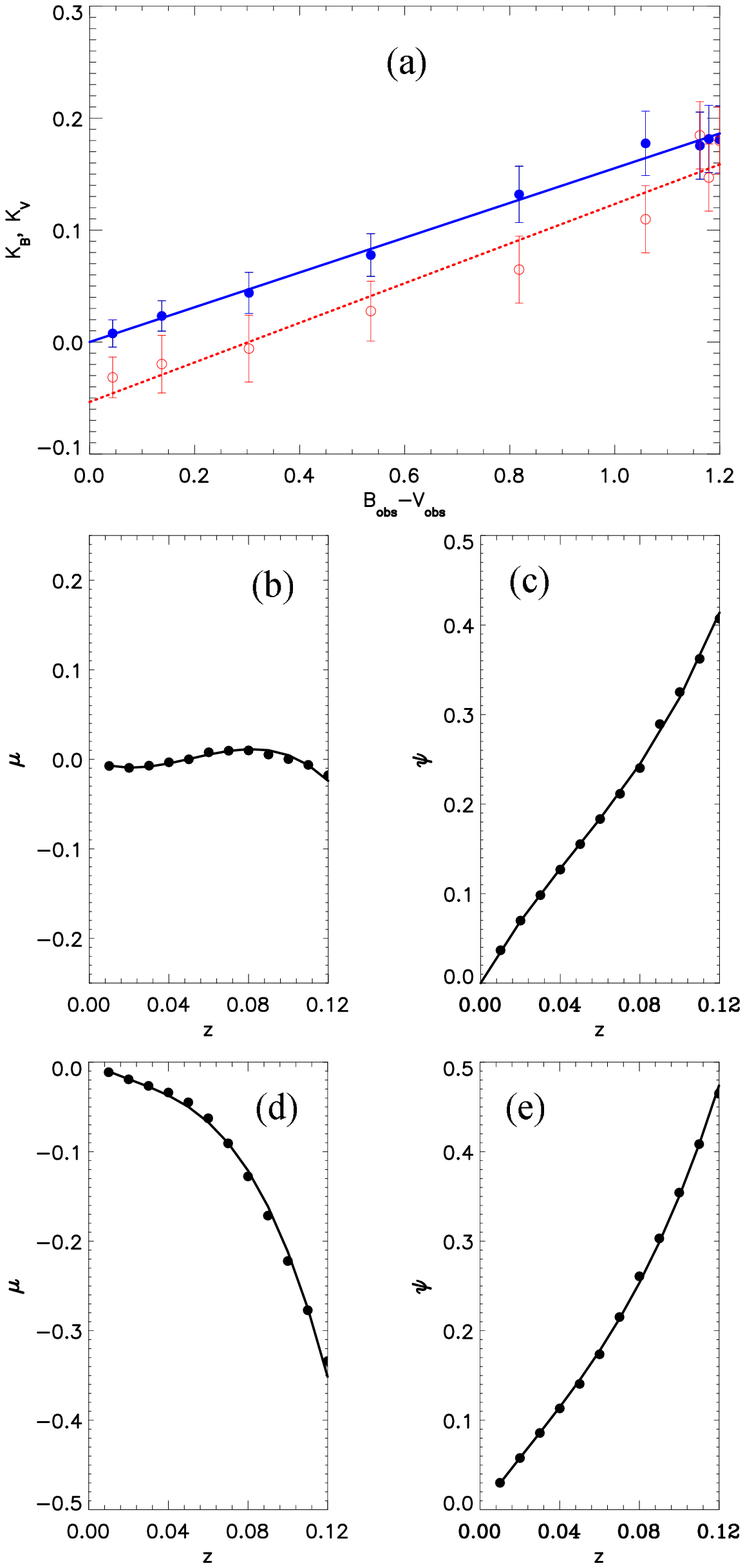}
\caption{Illustration of $K$ correction method.  ({\it a}) $K$ corrections at $z = 0.05$ to $B$ magnitudes (full circles) and $V$ magnitudes (open circles) of SN1994D {\it vs.}\ observed $\bv$ color.  The lines are linear fits with intercept $\mu$ and slope $\psi$.  ({\it b-e}) Dependence on $z$ of ({\it b,d}) $\mu$ and ({\it c,e}) $\psi$ for ({\it b,c}) $B$ and ({\it d,e}) $V$ magnitudes.  The fit curves are cubic polynomials in $z$.} \label{Kdetails}
\end{figure}


Additional details of the $K$ correction method are provided in Fig.~\ref{Kdetails} and Table \ref{Kcoeffs}.  Figure \ref{Kdetails}(a) exhibits linear fits to the dependence upon $\bv$ color of the $K$ corrections to $B$ and $V$ magnitudes of SN1994D.  The $1 \sigma$ error bars are estimated by varying the choice of library spectral template while warping it to reproduce the observed SN color.  The slopes and intercepts of these lines are fit to cubic functions of $z$ shown in Figs.~\ref{Kdetails}(b-e), whose fit coefficients are provided in Table \ref{Kcoeffs}.  The $K$ corrections to $B_{\rm max}$, $V_{\rm max}$, $\Delta m_{15}$, $B_{BV}$, and $\beta_{BV}$ are obtained by evaluating the $K$ corrections to $B$ and/or $V$ magnitudes at the appropriate $\bv$ colors. 


\tabletypesize{\footnotesize}
\setlength{\tabcolsep}{4pt}
\begin{deluxetable}{crrr}
\tablewidth{0pt}
\tablecolumns{4}
\tableheadfrac{}
\tablecaption{Coefficients of $z$, $z^2$, and $z^3$ from the cubic fits shown in Fig.~\ref{Kdetails}({\it b}), ({\it c}), ({\it d}), and ({\it e}).\label{Kcoeffs}}
\tabletypesize{}
\tablehead{\colhead{\begin{tabular}{c} Fig.\\ \ref{Kdetails} \end{tabular}}
&\colhead{\begin{tabular}{c} Coefficient \\ of $z$   \end{tabular}}
&\colhead{\begin{tabular}{c} Coefficient \\ of $z^2$ \end{tabular}}
&\colhead{\begin{tabular}{c} Coefficient \\ of $z^3$ \end{tabular}}
}
\startdata
({\it b}) &$ -0.97119\phn  $&$  28.9034\phn  $&$ -187.314\phn $  \\
({\it c}) &$  3.86474\phn  $&$ -23.3410\phn  $&$  165.731\phn $  \\
({\it d}) &$ -1.17487\phn  $&$  16.5470\phn  $&$ -259.775\phn $  \\
({\it e}) &$  3.04291\phn  $&$ -10.4068\phn  $&$  149.714\phn $  \\
\enddata
\end{deluxetable}


\section{NOISE DETAILS} \label{appN}

In gaussian approximation and in the basis $\{B_{\rm max},B_{BV}\}$ of the measured $B$ magnitudes, the noise in a single SN's measured luminosity from all sources other than photometric error is expressed by the noise covariance matrix 
\begin{equation}
Y = \pmatrix{\sigma^2_{\rm max} & V_c \cr
             V_c   & \sigma^2_{BV} \cr } \; , \label{eqM}
\end{equation} 
where the off-diagonal element $V_c$ in principle is bounded only by $-\sigma_{\rm max} \sigma_{BV} \le V_c \le \sigma_{\rm max} \sigma_{BV}$.  

Because allocating the measured noise between $B_{\rm max}$ and $B_{BV}$ is challenging statistically, it is advantageous to approximate $Y$ as the sum of two simpler matrices
\begin{equation}
Y = \pmatrix{\sigma^2_{u,{\rm max}} & 0 \cr
             0   & \sigma^2_{u,BV}  } +
    \pmatrix{V_c & V_c \cr
             V_c & V_c \cr } \; . \label{eqN}
\end{equation} 
The first covariance matrix arises from sources of noise $\sigma^2_{u,\dots}$ that are completely uncorrelated between $B_{\rm max}$ and $B_{BV}$, while the second represents noise sources (including peculiar velocity and, at higher redshift than considered here, weak lensing) that affect both measurements equally.  Equation (\ref{eqN}) is an approximation to equation (\ref{eqM}) only in that it restricts the off-diagonal element of $Y$ to the range  $0 \le V_c \le \min{(\sigma^2_{\rm max},\sigma^2_{BV})}$.  (This is the ``mild assumption'' to which reference is made in \S \ref{noiseProp}.)  Intrinsic noise can contribute to either of the terms in equation (\ref{eqN}).  

After the noise covariance matrix $Y$ from equation (\ref{eqN}) is transformed to the basis $\{\Delta B, \langle B \rangle\}$, its diagonal elements are
\begin{mathletters}
\begin{eqnarray}
\sigma^2_{\Delta B} &=& \sigma^2_{u,{\rm max}} + \sigma^2_{u,BV} 
  \; ; \label{eqO}\\
\sigma^2_{\langle B \rangle} &=& {\textstyle{\frac{1}{4}}}
  (\sigma^2_{u,{\rm max}} + \sigma^2_{u,BV}) + 
  V_c 
  \; . \label{eqP}
\end{eqnarray}
\end{mathletters}
Apart from sources of noise $V_c$ that are common to both $B_{\rm max}$ and $B_{BV}$, it is evident from eqs.~(\ref{eqO}-\ref{eqP}) that the rms noise $n_{\Delta B}$ that is measured in the comparison of $B_{\rm max}$ to $B_{BV}$ propagates into $\langle B \rangle$ as $n_{\langle B \rangle} = n_{\Delta B}/2$.

In \S \ref{noiseProp}, reference is made also to the possibility of using the redshift dependence of the common-mode noise to resolve the effects of peculiar velocity from those of other sources.  Such analysis, which is not pursued here due to insufficient statistics, would use an equation of the form
\begin{equation}
\sigma^2_{\rm max} = \sigma^2_{u,{\rm max}} + V_c^{\rm int}
 + \left( {{C} \over {z}} \, {{v_{\rm pec}} \over {c}} \right)^2 \; , \label{eqPQ}
\end{equation}
where $V_c^{\rm int}$ is the portion of $V_c$ that is $z$-independent.  The factor $C(z)$ is equal to 5, multiplied by a correction that, for a flat universe with $\Omega_M = 0.3$, ranges from 1.00 to 1.05 over the redshift range (0.003 to 0.079) of the actual SNe in our fitted sample.

\section{COLOR RESOLUTION DETAILS} \label{appR}

In \S \ref{colorHub}, reference is made to two different methods of estimating the correction to ${\cal R}$ from the effects of finite color resolution.  In these methods the average color $\langle E \rangle$ (equation (\ref{eqI})) was used.  In the first method, the SNe were redistributed in color so that a twice-resolution-smeared color ideogram of the redistributed SNe matched a once-resolution-smeared ideogram of the unredistributed SNe.  During the redistribution, the SN excursions were limited by requiring $\chi^2/{\rm dof} = 1$, {\it i.e.}~the rms excursion of a  SN was limited to one standard deviation in its own color resolution.  Using the color-redistributed SNe as indicators of the true SN color distribution, a linear least-squares fit was made to a scatter plot of redistributed color {\it vs.}~observed color for each SN.  The fit slope was 0.962, implying that when ${\cal R}$ is fit to the observed SN color it is underestimated by 3.8\%.

In the second method, a once-resolution-smeared color ideogram first was mildly smoothed using a multigaussian to yield an ``apparent'' color distribution.  Next, a continuous ``true'' color distribution was identified such that after two gaussian smearings it matched the apparent distribution.  In this method the smearing width was set to the median color resolution of the SN set.  Next, by smearing the true distribution only once, the average true color that contributes to each apparent color was determined.  Then an analytic function was fitted to a plot of (average true color $-$ apparent color) {\it vs.}~apparent color.  Using this function, the ratio $\rho$ of average true color to apparent color was constructed as a function of apparent color.  Finally, the global fit described in \S \ref{Bavg} was repeated with each SN's observed color multiplied by $\rho$.  Relative to the standard fit value, ${\cal R}$ was found to increase by 7.0\%.

\end{appendix}

{}

\end{document}